\documentclass[10pt, conference, letterpaper]{IEEEtran}
\IEEEoverridecommandlockouts
\usepackage{cite}
\usepackage{amsmath,amssymb,amsfonts}
\usepackage{algorithmic,algorithm}
\usepackage{graphicx}
\usepackage{caption}
\usepackage{subcaption}

\usepackage{cleveref}
\usepackage{textcomp}
\usepackage{xcolor}
\usepackage{bbm}
\usepackage{tikz}
\usepackage{color,soul}
\usetikzlibrary{patterns}

\newtheorem{theorem}{Theorem}
\newtheorem{lemma}{Lemma}
\newtheorem{proposition}{Proposition}
\newtheorem{definition}{Definition}
\newtheorem{corollary}{Corollary}
\newtheorem{remark}{Remark}

\newcommand{\av}[1]{\overline{#1}}
\newcommand{\aaoi}{\overline{\Delta}}
\newcommand{\valpha}{\boldsymbol{\alpha}}

\def\BibTeX{{\rm B\kern-.05em{\sc i\kern-.025em b}\kern-.08em
    T\kern-.1667em\lower.7ex\hbox{E}\kern-.125emX}}
\begin{document}

\title{3-Competitive Policy for Minimizing Age of Information in Multi-Source M/G/1 Queuing Model
\thanks{We acknowledge support of the Department of Atomic Energy, Government of India, under project no. RTI4001.}
}

\author{\IEEEauthorblockN{Kumar Saurav}
\IEEEauthorblockA{\textit{School of Technology and Computer Science} \\
\textit{Tata Institute of Fundamental Research}\\
Mumbai, India. \\
kumar.saurav@tifr.res.in}
}
\vspace{-1in}
\def\onehalf{\frac{1}{2}}
\def\etal{et.\/ al.\/}
\newcommand{\bydef}{\triangleq}
\newcommand{\tr}{{\it{tr}}}
\def\SNR{{\textsf{SNR}}}
\def\bydef{:=}
\def\bba{{\mathbb{a}}}
\def\bbb{{\mathbb{b}}}
\def\bbc{{\mathbb{c}}}
\def\bbd{{\mathbb{d}}}
\def\bbee{{\mathbb{e}}}
\def\bbff{{\mathbb{f}}}
\def\bbg{{\mathbb{g}}}
\def\bbh{{\mathbb{h}}}
\def\bbi{{\mathbb{i}}}
\def\bbj{{\mathbb{j}}}
\def\bbk{{\mathbb{k}}}
\def\bbl{{\mathbb{l}}}
\def\bbm{{\mathbb{m}}}
\def\bbn{{\mathbb{n}}}
\def\bbo{{\mathbb{o}}}
\def\bbp{{\mathbb{p}}}
\def\bbq{{\mathbb{q}}}
\def\bbr{{\mathbb{r}}}
\def\bbs{{\mathbb{s}}}
\def\bbt{{\mathbb{t}}}
\def\bbu{{\mathbb{u}}}
\def\bbv{{\mathbb{v}}}
\def\bbw{{\mathbb{w}}}
\def\bbx{{\mathbb{x}}}
\def\bby{{\mathbb{y}}}
\def\bbz{{\mathbb{z}}}
\def\bb0{{\mathbb{0}}}

\def\bydef{:=}
\def\ba{{\mathbf{a}}}
\def\bb{{\mathbf{b}}}
\def\bc{{\mathbf{c}}}
\def\bd{{\mathbf{d}}}
\def\bee{{\mathbf{e}}}
\def\bff{{\mathbf{f}}}
\def\bg{{\mathbf{g}}}
\def\bh{{\mathbf{h}}}
\def\bi{{\mathbf{i}}}
\def\bj{{\mathbf{j}}}
\def\bk{{\mathbf{k}}}
\def\bl{{\mathbf{l}}}
\def\bm{{\mathbf{m}}}
\def\bn{{\mathbf{n}}}
\def\bo{{\mathbf{o}}}
\def\bp{{\mathbf{p}}}
\def\bq{{\mathbf{q}}}
\def\br{{\mathbf{r}}}
\def\bs{{\mathbf{s}}}
\def\bt{{\mathbf{t}}}
\def\bu{{\mathbf{u}}}
\def\bv{{\mathbf{v}}}
\def\bw{{\mathbf{w}}}
\def\bx{{\mathbf{x}}}
\def\by{{\mathbf{y}}}
\def\bz{{\mathbf{z}}}
\def\b0{{\mathbf{0}}}
\def\opt{\mathsf{OPT}}
\def\on{\mathsf{ON}}
\def\off{\mathsf{OF}}
\def\bA{{\mathbf{A}}}
\def\bB{{\mathbf{B}}}
\def\bC{{\mathbf{C}}}
\def\bD{{\mathbf{D}}}
\def\bE{{\mathbf{E}}}
\def\bF{{\mathbf{F}}}
\def\bG{{\mathbf{G}}}
\def\bH{{\mathbf{H}}}
\def\bI{{\mathbf{I}}}
\def\bJ{{\mathbf{J}}}
\def\bK{{\mathbf{K}}}
\def\bL{{\mathbf{L}}}
\def\bM{{\mathbf{M}}}
\def\bN{{\mathbf{N}}}
\def\bO{{\mathbf{O}}}
\def\bP{{\mathbf{P}}}
\def\bQ{{\mathbf{Q}}}
\def\bR{{\mathbf{R}}}
\def\bS{{\mathbf{S}}}
\def\bT{{\mathbf{T}}}
\def\bU{{\mathbf{U}}}
\def\bV{{\mathbf{V}}}
\def\bW{{\mathbf{W}}}
\def\bX{{\mathbf{X}}}
\def\bY{{\mathbf{Y}}}
\def\bZ{{\mathbf{Z}}}
\def\b1{{\mathbf{1}}}

\def\bbA{{\mathbb{A}}}
\def\bbB{{\mathbb{B}}}
\def\bbC{{\mathbb{C}}}
\def\bbD{{\mathbb{D}}}
\def\bbE{{\mathbb{E}}}
\def\bbF{{\mathbb{F}}}
\def\bbG{{\mathbb{G}}}
\def\bbH{{\mathbb{H}}}
\def\bbI{{\mathbb{I}}}
\def\bbJ{{\mathbb{J}}}
\def\bbK{{\mathbb{K}}}
\def\bbL{{\mathbb{L}}}
\def\bbM{{\mathbb{M}}}
\def\bbN{{\mathbb{N}}}
\def\bbO{{\mathbb{O}}}
\def\bbP{{\mathbb{P}}}
\def\bbQ{{\mathbb{Q}}}
\def\bbR{{\mathbb{R}}}
\def\bbS{{\mathbb{S}}}
\def\bbT{{\mathbb{T}}}
\def\bbU{{\mathbb{U}}}
\def\bbV{{\mathbb{V}}}
\def\bbW{{\mathbb{W}}}
\def\bbX{{\mathbb{X}}}
\def\bbY{{\mathbb{Y}}}
\def\bbZ{{\mathbb{Z}}}

\def\cA{\mathcal{A}}
\def\cB{\mathcal{B}}
\def\cC{\mathcal{C}}
\def\cD{\mathcal{D}}
\def\cE{\mathcal{E}}
\def\cF{\mathcal{F}}
\def\cG{\mathcal{G}}
\def\cH{\mathcal{H}}
\def\cI{\mathcal{I}}
\def\cJ{\mathcal{J}}
\def\cK{\mathcal{K}}
\def\cL{\mathcal{L}}
\def\cM{\mathcal{M}}
\def\cN{\mathcal{N}}
\def\cO{\mathcal{O}}
\def\cP{\mathcal{P}}
\def\cQ{\mathcal{Q}}
\def\cR{\mathcal{R}}
\def\cS{\mathcal{S}}
\def\cT{\mathcal{T}}
\def\cU{\mathcal{U}}
\def\cV{\mathcal{V}}
\def\cW{\mathcal{W}}
\def\cX{\mathcal{X}}
\def\cY{\mathcal{Y}}
\def\cZ{\mathcal{Z}}

\def\sfA{\mathsf{A}}
\def\sfB{\mathsf{B}}
\def\sfC{\mathsf{C}}
\def\sfD{\mathsf{D}}
\def\sfE{\mathsf{E}}
\def\sfF{\mathsf{F}}
\def\sfG{\mathsf{G}}
\def\sfH{\mathsf{H}}
\def\sfI{\mathsf{I}}
\def\sfJ{\mathsf{J}}
\def\sfK{\mathsf{K}}
\def\sfL{\mathsf{L}}
\def\sfM{\mathsf{M}}
\def\sfN{\mathsf{N}}
\def\sfO{\mathsf{O}}
\def\sfP{\mathsf{P}}
\def\sfQ{\mathsf{Q}}
\def\sfR{\mathsf{R}}
\def\sfS{\mathsf{S}}
\def\sfT{\mathsf{T}}
\def\sfU{\mathsf{U}}
\def\sfV{\mathsf{V}}
\def\sfW{\mathsf{W}}
\def\sfX{\mathsf{X}}
\def\sfY{\mathsf{Y}}
\def\sfZ{\mathsf{Z}}

\def\bydef{:=}
\def\sfa{{\mathsf{a}}}
\def\sfb{{\mathsf{b}}}
\def\sfc{{\mathsf{c}}}
\def\sfd{{\mathsf{d}}}
\def\sfee{{\mathsf{e}}}
\def\sfff{{\mathsf{f}}}
\def\sfg{{\mathsf{g}}}
\def\sfh{{\mathsf{h}}}
\def\sfi{{\mathsf{i}}}
\def\sfj{{\mathsf{j}}}
\def\sfk{{\mathsf{k}}}
\def\sfl{{\mathsf{l}}}
\def\sfm{{\mathsf{m}}}
\def\sfn{{\mathsf{n}}}
\def\sfo{{\mathsf{o}}}
\def\sfp{{\mathsf{p}}}
\def\sfq{{\mathsf{q}}}
\def\sfr{{\mathsf{r}}}
\def\sfs{{\mathsf{s}}}
\def\sft{{\mathsf{t}}}
\def\sfu{{\mathsf{u}}}
\def\sfv{{\mathsf{v}}}
\def\sfw{{\mathsf{w}}}
\def\sfx{{\mathsf{x}}}
\def\sfy{{\mathsf{y}}}
\def\sfz{{\mathsf{z}}}
\def\sf0{{\mathsf{0}}}

\def\Nt{{N_t}}
\def\Nr{{N_r}}
\def\Ne{{N_e}}
\def\Ns{{N_s}}
\def\Es{{E_s}}
\def\No{{N_o}}
\def\sinc{\mathrm{sinc}}
\def\dmin{d^2_{\mathrm{min}}}
\def\vec{\mathrm{vec}~}
\def\kron{\otimes}
\def\Pe{{P_e}}
\newcommand{\expeq}{\stackrel{.}{=}}
\newcommand{\expg}{\stackrel{.}{\ge}}
\newcommand{\expl}{\stackrel{.}{\le}}
\def\SIR{{\mathsf{SIR}}}

\def\nn{\nonumber}

\maketitle
\vspace{-0.1in}
\begin{abstract}
	We consider a multi-source network with a common monitor, where fresh updates are generated at each source, following a Poisson process. At any time, at most one source can transmit its update to the monitor, and transmission time for updates of each source follows some general distribution. The goal is to find a causal scheduling policy such that at any time, the latest update available at each source is fresh. In this paper, we quantify freshness using the age of information (AoI) metric, and propose a randomized policy, which we show is 3-competitive with respect to Pareto-optimal policies (that minimize the expected average AoI of each source). We also show that for a particular choice of the randomization parameter, the proposed randomized policy is 3-competitive with respect to an optimal policy that minimizes the weighted sum of the expected average AoI of all sources.
\end{abstract}


\section{Introduction} \label{sec:intro}



Reliance on time-sensitive networked applications (such as remote monitoring, telehealth services, control, etc.) for critical roles, have necessitated the need for a general transmission policy that could ensure timely delivery of fresh updates of each source, at the corresponding destination. The policy should mitigate the effect of constraints on update generation, transmission delay, capacity of the shared channel, number of sources, etc., and must be simple to implement. 
In this paper, we propose a particular randomized policy, and show that it adheres to the above-mentioned requirements.

In particular, we consider a general multi-source M/G/1 queuing model, where 
at each source, the updates are generated with exponentially distributed inter-generation time. Also, at any time, at most one source can transmit, and transmission delay for each source follows some general distribution. 
Note that the mean update generation rate and transmission delay distribution may be different for each source, however they do not change with time.

The objective is to find an online transmission policy (in short, policy/online policy), such that at any time, the latest update of each source available at the destination is fresh. Formally, we quantify freshness using \emph{age of information (AoI)} metric \cite{kaul2012real,kaul2012status,saurav2021game,saurav2021online}. At any time, AoI of a source is equal to the time elapsed since the generation time of the latest update of the source, available at the monitor. Thus, for any policy, if the long-term \emph{average AoI (AAoI)} is small (less than a certain threshold) for each source, then the policy is said to have achieved timely delivery of fresh updates. In the considered model, 
this threshold $\alpha_\ell$ is provided for each source $\ell$. 

Ideally, for each source $\ell$, its AAoI should be less than 
$\alpha_\ell$. However, due to constraints on availability of updates, transmission delays, large number of sources, etc., this may not be possible for arbitrary vector $\valpha=(\alpha_1,\alpha_2,\cdots)$, under any policy. 
Also, the set $\cC$ of vectors $\valpha$ 
for which this is possible under some online policy $\pi^\star$, is not known. 
Therefore, in this paper, we first derive a necessary condition to characterize this set $\cC$. 
Then, we propose a simple randomized policy $\pi_R$, that at any time, among all the sources, picks a source $\ell$ with a fixed probability $p_\ell$, and transmits its update (if it has an update). We show that for any vector $\valpha$ that satisfies the derived necessary condition, 
the expected AAoI of each source $\ell$ under $\pi_R$ is at most $3\alpha_\ell$. 

In the later part of this paper, we also consider the setting, where instead of threshold vector $\valpha$, only relative weights $\bw=(w_1,w_2,\cdots)$ are known for the sources, and the objective is to minimize the weighted sum of the expected AAoI (WSAAoI) of the sources, following an online policy. We show that for a particular choice of randomization parameter $p_\ell$ (corresponding to each source $\ell$), the same randomized policy $\pi_R$ proposed for the earlier setting, 
guarantees WSAAoI that is at most three times the WSAAoI for an optimal online policy $\pi^\star$ (that has minimum WSAAoI).

Although the theoretical guarantee for $\pi_R$ has a multiplicative gap of $3$ (relative to $\pi^\star$), 
for the general multi-source M/G/1 queuing model, this is a significant result. In prior work, for multi-source setting with stochastic packet generation, the best guarantee for any online policy has a multiplicative gap of $4$  \cite{kadota2019minimizing} (the results in \cite{kadota2019minimizing} are for slotted time M/M/1 queuing model, unlike the general continuous time M/G/1 model considered in this paper). 

In most of the prior work on AoI (e.g., \cite{kadota2018optimizing,kadota2018scheduling,kadota2019minimizing,saurav2021minimizing}), a major reason for the large gap in AAoI guarantee for an online policy with respect to $\pi^\star$, is the use of a weak lower bound on the AAoI of $\pi^\star$. Generally (as in \cite{kadota2018optimizing,kadota2018scheduling,kadota2019minimizing,saurav2021minimizing,bhat2020throughput}), the lower bound disregards the effect of variance $\sigma_\ell^2$ of inter-generation time of updates on the AAoI of sources under $\pi^\star$. 
In this paper, we derive a better lower bound that accounts for the effect of variance $\sigma_\ell^2$ on the AAoI of $\pi^\star$ (at least partially), and hence, are able to minimize the gap in the guarantee, relative to prior work.

Currently, in this paper, the major limitation in improving the theoretical guarantee for $\pi_R$ (or, designing a better online policy), is the weak lower bound on the waiting time of updates under $\pi^\star$. For any transmitted update, the waiting time is equal to the difference between the time when the update got generated, and time when a policy transmits the update. Thus, waiting times have a significant impact on the AAoI of any policy. 
However, for general multi-source setup with stochastic update generation, to the best of our knowledge, none of the prior work has been able to derive a lower bound on the waiting time of updates under $\pi^\star$, that is better than $0$. Hence, improving the lower bound on the waiting times of $\pi^\star$, remains an active problem for our future work. 

Meanwhile, under specific condition, such as when number of sources is $1$ \cite{sun2017update,saurav2021minimizing}, or when fresh updates are available at all times \cite{kadota2018optimizing,kadota2018scheduling,sun2017update}, better lower bounds are known. In fact, when number of sources is $1$, \cite{sun2017update,saurav2021minimizing} proposed online policies that where shown to be theoretically optimal. 
Hence, for fair evaluation of the performance of $\pi_R$ in these settings, we compared $\pi_R$ directly with $\pi^\star$ (instead on lower bound on AAoI) using numerical simulations. 
In particular, we considered the setting of \cite{sun2017update} (a single source that can generate a new update at any time, and the transmission delay for each update follows some general distribution), 
and found that despite the simplicity, and generality of $\pi_R$, 
its AAoI is close to the AAoI of $\pi^\star$ (for the two transmission delay distributions that we considered). 

The rest of this paper is organised as follows. In Section \ref{sec:SysModel}, we discuss the considered M/G/1 queuing model in detail, and formally define the objective. 
In Section \ref{sec:CapacityRegion}, we derive the necessary condition to characterize the set $\cC$ of AAoI threshold vector $\valpha$ that an optimal online policy may achieve. In Section \ref{sec:RandomizedPolicy}, we propose the randomized policy $\pi_R$, and derive an upper bound on the expected AAoI of each source under $\pi_R$. We show that for any $\valpha\in\cC$, the expected AAoI of the sources under $\pi_R$ is at most $3\valpha$. In Section \ref{sec:AoISumMin}, we consider weighted sum expected AAoI (i.e., WSAAoI) minimization problem, and generalize $\pi_R$ (and the corresponding guarantee) for this setting. Finally, in Section \ref{sec:NumericalResults}, we discuss the numerical simulation results.

\section{System Model} \label{sec:SysModel}

Consider a system consisting of $N$ sources and a monitor. At each source $\ell\in\{1,\cdots,N\}$, updates (henceforth, \emph{packets}) are generated with exponentially distributed inter-generation time $X_\ell$, with mean $\mu_\ell<\infty$. The sources transmit their packets to the monitor, over a common channel, that at any time $t$, allows at most one source to transmit (one packet). 
Each packet transmitted by source $\ell$ gets received at the monitor after random transmission delay $d_\ell\sim\cD_\ell$, where $\cD_\ell$ is some general distribution with mean $\gamma_\ell<\infty$.\footnote{For different sources, $\cD_\ell$'s may belong to different family of distributions.} 
The sources may choose whether to transmit a packet, 
or discard it, but only until the transmission of the packet is initiated. Once initiated, a transmission cannot be preempted. 

\begin{definition} \label{def:channel-busy}
	While a packet is under transmission, the channel is said to be busy. Otherwise, the channel is free. A transmission can be initiated only when the channel is free.
\end{definition}

\begin{figure}
	\begin{center}
		\begin{tikzpicture}[thick,scale=0.8, every node/.style={scale=1}]
		\draw[->] (-0.25,0) to (7.8,0) node[below]{time ($t$)};
		\draw[->] (0.35,-0.1) node[below]{$g_{\ell 0}$} to (0.35,2.9) node[below left]{$\Delta_{\ell}(t)$};
		\draw (0.35,0) to (2.7,2.35) to (2.7,0.7) to (4.8,2.8) to (4.8,0.8) to (6.4,2.4) to (6.4,0.65) to (6.75,1); 
		
		
		\draw[loosely dotted] (7.0,1) to (7.7,1); 
		
		\draw (2,-0.1) node[below]{$g_{\ell 1}$} to (2,0.1);
		\draw (4,-0.1) node[below]{$g_{\ell 2}$} to (4,0.1);
		\draw (5.75,-0.1) node[below]{$g_{\ell 3}$} to (5.75,0.1); 
		
		\draw (2.7,-0.1) node[below]{$r_{\ell 1}$} to (2.7,0.1);
		\draw (4.8,-0.1) node[below]{$r_{\ell 2}$} to (4.8,0.1);
		\draw (6.4,-0.1) node[below]{$r_{\ell 3}$} to (6.4,0.1); 
		
		
		\draw[dashed] (2,0) to (2.7,0.7) to (2.7,0.1);
		\draw[dashed] (4,0) to (4.8,0.8) to (4.8,0.1);
		\draw[dashed] (5.75,0) to (6.4,0.65) to (6.4,0.1);
		
		\draw[|<->] (0.35,-1) -- (2,-1) node[rectangle,inner sep=-1pt,midway,fill=white]{$T_{\ell 1}$}; 
		\draw[|<->] (2,-1) -- (4,-1) node[rectangle,inner sep=-1pt,midway,fill=white]{$T_{\ell 2}$};
		\draw[|<->|<] (4,-1) -- (5.8,-1) node[rectangle,inner sep=-1pt,midway,fill=white]{$T_{\ell 3}$};
		
		
		\draw[loosely dotted] (6,-1) to (6.8,-1);
		
		\end{tikzpicture}
		\caption{Sample AoI plot for source $\ell$ (assuming AoI at time $g_{\ell 0}$ to be $0$). 
		Here, $g_{\ell i}$ and $r_{\ell i}$ respectively denote the generation time and transmission completion time of packet $i\ge 1$. Also, $T_{\ell i}$ ($\forall i\ge1$) denote the inter-generation time of transmitted packets of source $\ell$. \vspace{-2ex}}
		\label{fig:multi-node-general-age} 
	\end{center}
\end{figure}
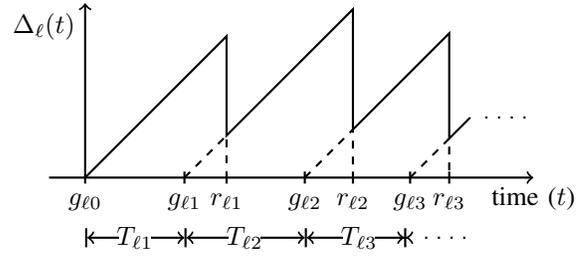

At any time $t$, the age of information (AoI) of a source $\ell$ (denoted $\Delta_\ell(t)$) is equal to the time elapsed since the generation time of the latest packet of the source that has been received at the monitor. Thus, as shown in Figure \ref{fig:multi-node-general-age}, $\Delta_\ell(t)=t-\lambda_\ell(t)$, where $\lambda_\ell(t)$ denotes the generation time of the latest update of source $\ell$ that has been received at the monitor until time $t$. Average AoI (in short, AAoI) of source $\ell$ until time $t$ is defined as
\begin{align} \label{eq:aaoi}
	\aaoi_\ell(t)=\frac{\int_{0}^{t}\Delta_\ell(i)di}{t}.
\end{align}

\begin{definition} \label{def:policy}
	A centralized online transmission policy (in short, a \emph{policy}) is an algorithm, that at each time $t$ (when the channel is free), using only the causal information available at all the sources at time $t$, decides which source gets to transmit (at time $t$). 
	In this paper, we only consider the set of policies $\Pi$ that never preempt any packet that is under transmission. 
\end{definition} 

The objective in this paper is to find a policy $\pi\in\Pi$, 
such that for any given vector $\valpha=(\alpha_1,\cdots,\alpha_N)$ of recommended AAoI for the $N$ sources, the policy $\pi$ minimizes the ratio of the long term AAoI of source $\ell$ to $\alpha_\ell$,  
To make this formal, for any given $\valpha$, we define the cost for a policy $\pi$ as 
\begin{align} \label{eq:cost}
	\Gamma(\pi;\valpha)=\max_\ell\left\{\bbE_\pi\left[\lim_{t\to\infty}\aaoi_\ell^\pi(t)\right]/\alpha_\ell\right\},
\end{align}
where $\bbE_\pi[\cdot]$ denotes expectation with respect to policy $\pi$, as well as the packet generation and transmission delay distribution of the sources. Thus, the objective is to find
\begin{align} \label{eq:objective}
	\pi^\star=\underset{\pi\in\Pi}{\arg\min}\ \ \Gamma(\pi;\valpha).
\end{align}

\begin{remark} \label{remark:PiON}
Since $\mu_\ell$ and $\gamma_\ell$ are finite for each source $\ell$, as shown in Appendix \ref{appendix:non-empty-PiON}, there exists a policy $\pi\in\Pi$, such that the expected AAoI of each source under policy $\pi$ is finite  (except when $N\to\infty$, in which case, $\Gamma(\pi;\valpha)\to\infty$ for all policies in $\Pi$, and hence, the objective \eqref{eq:objective} becomes meaningless). 
Hence, the expected AAoI of each source under $\pi^\star$ must also be finite. Therefore, without loss of generality, in the rest of this paper, we disregard the policies in $\Pi$ for which the expected AAoI of any of the source is infinity, and assume that for any policy $\pi\in\Pi$, the expected AAoI of all the sources is finite. 
\end{remark}

Note that $\Gamma(\pi;\valpha)\le1$ implies that the AAoI of  each source $\ell$ under policy $\pi$ is at most $\alpha_\ell$. However, for arbitrary $\valpha$, such policy $\pi$ may not exist (e.g., if $\valpha=(0,\cdots,0)$, $\Gamma(\pi,\valpha)=\infty$, $\forall\pi\in\Pi$). Therefore, for problem \eqref{eq:objective} to be meaningful, we only consider $\valpha\in\cC$, where 
\begin{align} \label{eq:capacity}
	\cC=\{\valpha\in\bbR^N \hspace{0.5ex} |\hspace{0.5ex} \exists\pi\in\Pi,\text{ s.t. } \Gamma(\pi;\valpha)\le 1\}
\end{align} 
denotes the set of feasible $\valpha$ (Definition \ref{def:feasible-alpha}), called the \emph{capacity region}. 
\begin{definition} \label{def:feasible-alpha}
	$\valpha$ is said to be feasible under policy $\pi\in\Pi$, if $\Gamma(\pi,\valpha)\le 1$, i.e., $\bbE_\pi[\lim_{t\to\infty}\aaoi_\ell(t)]\le\alpha_\ell$, $\forall \ell$.
\end{definition}

Further, because the cost $\Gamma(\pi;\valpha)$ \eqref{eq:cost} for a policy $\pi$ depends on $\valpha\in\cC$ (e.g., when $\valpha\to\infty$, $\Gamma(\pi;\valpha)\to 0$ for all reasonable policies in $\Pi$), we quantify the performance of policy $\pi$ using its competitive ratio $\textsc{CR}_\pi$ \eqref{eq:CR-def}, which is equal to the cost $\Gamma(\pi;\valpha)$ \eqref{eq:cost} for the policy, maximized over all $\valpha\in\cC$.
\begin{align} \label{eq:CR-def}
	\textsc{CR}_\pi=\underset{\valpha\in\cC}{\max} \hspace{1ex}\Gamma(\pi,\valpha)=\underset{\valpha\in\cC}{\max} \max_\ell\left\{\frac{\bbE_\pi[\underset{t\to\infty}{\lim}\aaoi_\ell(t)]}{\alpha_\ell}\right\}.
\end{align} 

To solve \eqref{eq:objective} and analyze \eqref{eq:CR-def}, it is critical to first characterize the capacity region $\cC$ \eqref{eq:capacity}. Hence, in next section, we derive a necessary condition that any $\valpha$ that lies in $\cC$, must satisfy.

\section{Capacity Region $\cC$} \label{sec:CapacityRegion} 

Consider the following lower bound on the expected AAoI of source $\ell$ under policy $\pi\in\Pi$.
\begin{lemma} \label{lemma:lower-bound}
	For any policy $\pi\in\Pi$, the expected AAoI of source $\ell$ satisfies 
	\begin{align} \label{eq:lower-bound}
		\bbE_\pi\left[\lim_{t\to\infty}\aaoi_\ell(t)\right]\ge \frac{1}{2}\left(\frac{\mu_\ell^2/2}{\bbE_\pi[\av{T}_\ell^\pi]}+\bbE_\pi[\av{T}_\ell^\pi]+2\gamma_\ell\right),
	\end{align}
    where $\bbE_\pi[\av{T}_\ell^\pi]$ denotes the expectation of the average of the inter-generation time of packets of source $\ell$ that are transmitted by policy $\pi$. Also, $\bbE_\pi[\av{T}_\ell^\pi]$ for each source $\ell$ satisfies 
    \begin{align} \label{eq:throughput-constraint}
    	\sum_{\ell=1}^N \frac{\gamma_\ell}{\bbE_\pi[\av{T}_\ell^\pi]}\le 1.
    \end{align}
\end{lemma}
\begin{IEEEproof}
	See Appendix \ref{appendix:proof-lemma-lb-ub}. 
\end{IEEEproof}
\begin{remark}
Note that $\mu_\ell^2$ is the variance of the exponentially distributed inter-generation time of packets (with mean $\mu_\ell$) at source $\ell$.
\end{remark}

Recall that $\valpha$ lies in $\cC$, only if it is feasible with respect to some policy $\pi\in\Pi$ (Definition \ref{def:feasible-alpha}), i.e., for some $\pi\in\Pi$, $\alpha_\ell\ge \bbE_\pi[\lim_{t\to\infty}\aaoi_\ell(t)]$, $\forall \ell\in\{1,\cdots,N\}$. 
Hence, using \eqref{eq:lower-bound}, we get that for any $\valpha\in\cC$, $\exists \pi\in\Pi$ such that for each source $\ell$,
\begin{align} \label{eq:NC-1}
	\alpha_\ell\ge 
	\frac{1}{2}\left(\frac{\mu_\ell^2/2}{\bbE_\pi[\av{T}_\ell^\pi]}+\bbE_\pi[\av{T}_\ell^\pi]+2\gamma_\ell\right).
\end{align}
But solving the quadratic inequality \eqref{eq:NC-1}, we find that \eqref{eq:NC-1} is true only if $\forall \ell\in\{1,\cdots,N\}$, $\alpha_\ell\ge\gamma_\ell$, $(\alpha_\ell-\gamma_\ell)^2\ge \mu_\ell^2/2$, and $\exists \pi\in\Pi$, such that for each source $\ell$,
\begin{align} \label{eq:NC-2}
	\bbE_\pi[\av{T}_\ell^\pi]\le (\alpha_\ell-\gamma_\ell)+\sqrt{(\alpha_\ell-\gamma_\ell)^2-\mu_\ell^2/2}. 
\end{align} 

Note that the conditions $\alpha_\ell>\gamma_\ell$ and $(\alpha_\ell-\gamma_\ell)^2\ge \mu_\ell^2/2$ are simultaneously true only if $\alpha_\ell\ge \gamma_\ell+\mu_\ell/\sqrt{2}$.
Also, \eqref{eq:throughput-constraint} and \eqref{eq:NC-2}, together imply $\sum_{\ell=1}^N\gamma_\ell/((\alpha_\ell-\gamma_\ell)+\sqrt{(\alpha_\ell-\gamma_\ell)^2-\mu_\ell^2/2})\le 1$. Hence, we get the following necessary condition for any $\valpha$ that lies in $\cC$. 
\begin{lemma} \label{lemma:NC}
	$\valpha$ lies in $\cC$, only if
	\begin{enumerate}
		\item $\alpha_\ell\ge \gamma_\ell+\mu_\ell/\sqrt{2}$, $\forall \ell\in\{1,\cdots,N\}$, and 
		\item $\sum_{\ell=1}^N\gamma_\ell/\av{T}_\ell^{\max}\le 1$, where 
		\begin{align} \label{eq:Tmax}
			\av{T}_\ell^{\max}=(\alpha_\ell-\gamma_\ell)+\sqrt{(\alpha_\ell-\gamma_\ell)^2-\mu_\ell^2/2}.
		\end{align}
	\end{enumerate}
\end{lemma} 

\begin{remark} \label{remark:Tmax-finite}
    For any source $\ell$, $\alpha_\ell\to\infty$ implies that the packets of source $\ell$ are not time-sensitive, and hence, need not be considered for the optimization problem \eqref{eq:objective}. Therefore, in the rest of this paper, without loss of generality, we assume that $\alpha_\ell$ is finite for each source $\ell\in\{1,\cdots,N\}$. 
    Thus, by definition, $\av{T}_\ell^{\max}$ \eqref{eq:Tmax} is also finite for each source $\ell$.
\end{remark}

\begin{corollary} \label{cor:lb-alpha}
	For any $\valpha\in\cC$, $\av{T}_\ell^{\max}$ \eqref{eq:Tmax} satisfies
	\begin{align} \label{eq:lb-alpha}
		\frac{1}{2}\left(\frac{\mu_\ell^2/2}{\av{T}_\ell^{\max}}+\av{T}_\ell^{\max}+2\gamma_\ell\right)=\alpha_\ell, \ \ \forall \ell\in\{1,\cdots,N\}.
	\end{align}
\end{corollary}
\begin{IEEEproof}
	Substituting \eqref{eq:Tmax} in the L.H.S. of \eqref{eq:lb-alpha}, we get  $\alpha_\ell$ (R.H.S. of \eqref{eq:lb-alpha}).
\end{IEEEproof}

Next, we propose a randomized policy $\pi_R\in\Pi$, and show that for any given $\valpha$ that satisfies the two conditions in Lemma \ref{lemma:NC}, $\pi_R$ has competitive ratio \eqref{eq:CR-def} at most $3$.

\section{Randomized Policy $\pi_R$} \label{sec:RandomizedPolicy}

Consider a randomized policy $\pi_R$ that at any time $t$, if the channel is free (Definition \ref{def:channel-busy}), among all the sources, picks source $\ell$, with probability
\begin{align} \label{eq:pick-prob}
	p_\ell=\frac{1/\av{T}_\ell^{\max}}{\sum_{i=1}^N 1/\av{T}_i^{\max}},
\end{align}
(where $\av{T}_\ell^{\max}$ is defined in \eqref{eq:Tmax}), and transmits its latest generated packet (if it has a 
packet to transmit, otherwise, idles for $d_\ell\sim\cD_\ell$ time units). If the channel is busy, $\pi_R$ waits for the channel to become free.

\begin{algorithm}
	\caption{Randomized Policy $\pi_R$.}
	\label{algo:threshold-policy}
	\begin{algorithmic}
		\STATE At any time $t$, 
		\IF{the channel is free}
		\STATE among $N$ sources, pick source $\ell$ with probability $p_\ell$ \eqref{eq:pick-prob}. 
		\IF{the source has a fresh packet to transmit}
		\STATE transmit its latest generated packet.
		\ELSE
		\STATE idle for $d_\ell\sim\cD_\ell$ time units.
		\ENDIF
		\ELSE
		\STATE wait for the channel to become free.
		\ENDIF
	\end{algorithmic}
\end{algorithm}

The following lemma provides an upper bound on the expected AAoI of each source $\ell$ under $\pi_R$.
\begin{lemma} \label{lemma:ub-piR}
	Under $\pi_R$ (Algorithm \ref{algo:threshold-policy}), the expected AAoI for each source $\ell$ satisfies 
	\begin{align} \label{eq:ub-piR}
		\bbE_R\left[\lim_{t\to\infty}\aaoi_\ell(t)\right]\le  \frac{1}{2}\left(\frac{\mu_\ell^2}{\av{T}_\ell^{\max}}+3\av{T}_\ell^{\max}+2\gamma_\ell\right).
	\end{align}
\end{lemma}
\begin{IEEEproof}
	See Appendix \ref{appendix:proof-lemma-lb-ub}. 
\end{IEEEproof}

The main result of this paper is as follows. 
\begin{theorem} \label{thm:CR-piR}
	The competitive ratio for $\pi_R$ (Algorithm \ref{algo:threshold-policy}) is $\textsc{CR}_{\pi_R}\le3$.
\end{theorem}
\begin{IEEEproof}
	From Corollary \ref{cor:lb-alpha} and Lemma \ref{lemma:ub-piR}, we get that for any $\valpha\in\cC$, $\bbE_R[\lim_{t\to\infty}\aaoi_\ell(t)]/\alpha_\ell\le 3$, for each source $\ell$. Hence, $\Gamma(\pi_R;\valpha)=\max_\ell\{\bbE_R[\lim_{t\to\infty}\aaoi_\ell(t)]/\alpha_\ell\}\le 3$. Thus, $\textsc{CR}_{\pi_R}=\underset{\valpha\in\cC}{\max}\ \ \Gamma(\pi_R;\valpha)\le 3$.
\end{IEEEproof}

Theorem \ref{thm:CR-piR} shows that if any policy can guarantee AAoI $\valpha$ for the sources, then under $\pi_R$, AAoI for the sources cannot be more than $3\valpha$. Given the ease in implementing $\pi_R$  ($\pi_R$ does not depend on the family of distribution that $\cD_\ell$'s belong to), this is an interesting result. However, in its current form, $\pi_R$ needs to know $\valpha$.
In next section, we generalize $\pi_R$ for systems where instead of $\valpha$, 
relative weights are known for the AAoI of the sources, 
and the objective is to minimize the weighted sum expected AAoI. We also derive an upper bound on the competitive ratio bound of $\pi_R$ with respect to an optimal policy that minimizes the weighted sum of the expected AAoI of all sources.

\section{Weighted Sum Expected AAoI Minimization} \label{sec:AoISumMin} 
Let $\bw=\{w_1,w_2,\cdots,w_N\}$ denote the relative weights for each source in the system, and define weighted sum expected AAoI to be
\begin{align}
	\Gamma(\pi;\bw)=\sum_{\ell=1}^N w_\ell\bbE_\pi\left[\lim_{t\to\infty}\aaoi_\ell(t)\right].
\end{align} 
The objective is to find an policy $\pi\in\Pi$ that minimizes $\Gamma(\pi;\bw)$ for any given $\bw$. Formally, the objective is to solve the following optimization problem:
\begin{align} \label{eq:pi-star}
	\pi^\star=\underset{\pi\in\Pi}{\arg\min}\ \ \Gamma(\pi;\bw).
\end{align}
Let $\valpha^\star=\{\alpha_1^\star,\cdots,\alpha_N^\star\}$ denote the expected AAoI for the source under $\pi^\star$ (as $t\to\infty$). Then, $\Gamma(\pi^\star;\bw)=\sum_{\ell=1}^N w_\ell\alpha_\ell^\star$. Using Lemma \ref{lemma:lower-bound}, we get  
\begin{align} \label{eq:lb-alpha-star-1}
	\sum_{\ell=1}^N w_\ell\alpha_\ell^\star\ge \frac{1}{2}\sum_{\ell=1}^N w_\ell\left(\frac{\mu_\ell^2/2}{\bbE_\star[\av{T}_\ell^\star]}+\bbE_\star[\av{T}_\ell^\star]+2\gamma_\ell\right),
\end{align}
where $\bbE_\star[\av{T}_{\ell}^\star]$ denotes the expected average inter-generation time of the packets of source $\ell$ that are transmitted under policy $\pi^\star$. Since $\pi^\star\in\Pi$, $\bbE_\star[\av{T}_\ell^\star]$ (for each source $\ell$) satisfy \eqref{eq:throughput-constraint}, and the R.H.S. of \eqref{eq:lb-alpha-star-1} is at least
\begin{align} \label{eq:lb-alpha-star-2}
	&\min_{1/\av{T}_\ell,\forall \ell}\ \ \frac{1}{2}\sum_{\ell=1}^N w_\ell\left(\frac{\mu_\ell^2/2}{\av{T}_\ell}+\av{T}_\ell+2\gamma_\ell\right), \\
	& \ \ \text{ s.t. } \sum_{\ell=1}^N (\gamma_\ell/\av{T}_\ell)\le1. \nonumber
\end{align}
Therefore, denoting the minimizer of \eqref{eq:lb-alpha-star-2} by  $1/\av{T}_\ell^o$, $\forall \ell$, we get $\sum_{\ell=1}^N w_\ell\alpha_\ell^\star\ge\sum_{\ell=1}^N w_\ell\alpha_\ell^o$, where 
\begin{align}
	\alpha_\ell^o=\frac{1}{2}\left(\frac{\mu_\ell^2/2}{\av{T}_\ell^o}+\av{T}_\ell^o+2\gamma_\ell\right)
\end{align}
 is a lower bound on $\alpha_\ell^\star$.
\begin{remark}
	Note that \eqref{eq:lb-alpha-star-2} is a convex optimization problem, and can be easily solved using standard optimization tools such as \emph{CVX} in Matlab. Hence, in the rest of this paper, we assume that $\av{T}_\ell^o$ and $\alpha_\ell^o$ are known for each source $\ell$.
\end{remark}

Now, for each source $\ell$, define 
\begin{align} \label{eq:p-ell-o}
	p_\ell^o=\frac{1/\av{T}_\ell^o}{\sum_{i=1}^N (1/\av{T}_i^o)}.
\end{align}

\begin{theorem} \label{thm:Sum-CR-piR}
	For the randomized policy $\pi_R$ (Algorithm \ref{algo:threshold-policy}) with $p_\ell=p_\ell^o$ \eqref{eq:p-ell-o} (for each source $\ell$), the weighted sum expected AAoI $\Gamma(\pi_R;\bw)\le 3\cdot\Gamma(\pi^\star;\bw)$, for any relative weight vector $\bw$.
\end{theorem}
\begin{IEEEproof}
	Replacing $\av{T}_\ell^{\max}$ by $\av{T}_{\ell}^o$ in \eqref{eq:lb-alpha}, \eqref{eq:pick-prob} and \eqref{eq:ub-piR}, we get 
	$p_\ell=p_\ell^o$ \eqref{eq:p-ell-o}, 
	\begin{gather} 
		\label{eq:alpha-o}
		\frac{1}{2}\left(\frac{\mu_\ell^2/2}{\av{T}_\ell^o}+\av{T}_\ell^o+2\gamma_\ell\right)=\alpha_\ell^o, \text{ and } \\ 
		\label{eq:ub-piR-o}
		\bbE_R\left[\lim_{t\to\infty}\aaoi_\ell(t)\right]\le  \frac{1}{2}\left(\frac{\mu_\ell^2}{\av{T}_\ell^o}+3\av{T}_\ell^o+2\gamma_\ell\right),
	\end{gather}
for each source $\ell$. Substituting \eqref{eq:alpha-o} in \eqref{eq:ub-piR-o}, and taking the weighted sum of the resulting expression over all sources, we get $\bbE_R[\lim_{t\to\infty}\aaoi_\ell(t)]\le \sum_{\ell=1}^N w_\ell (3\alpha_\ell^o)$. Since $\sum_{\ell=1}^N w_\ell \alpha_\ell^o\le \sum_{\ell=1}^N w_\ell\alpha_\ell^\star$, we get
	$$\Gamma(\pi_R;\bw)=\bbE_R\left[\lim_{t\to\infty}\aaoi_\ell(t)\right]\le 3\sum_{\ell=1}^N w_\ell \alpha_\ell^\star=3\cdot\Gamma(\pi^\star;\bw). \IEEEQEDhereeqn$$ 
\end{IEEEproof}

Theorem \ref{thm:Sum-CR-piR} shows that the weighted sum expected AAoI for the randomized policy $\pi_R$ is at most three times compared to any other policy in $\pi$. For M/G/1 queuing model, this is the best guarantee known so far.
Also, the proof of Theorem \ref{thm:Sum-CR-piR} provides a general recipe for generalizing the results for `per source AAoI minimization problem' to the `weighted sum AAoI minimization problem'. 



\section{Numerical Results} \label{sec:NumericalResults}

Theorems \ref{thm:CR-piR} and \ref{thm:Sum-CR-piR} provide some important analytical results regarding the performance of randomized policy $\pi_R$ (Algorithm \ref{algo:threshold-policy}). In this section, we use numerical simulations to verify these results, and derive new insights. 

\begin{remark}
For all the simulations, we assume the initial AoI of the sources to be $0$, and the time horizon $t=10^6$ units. 
\end{remark}

\begin{figure} 
	\centerline{\includegraphics[width=0.85\linewidth]{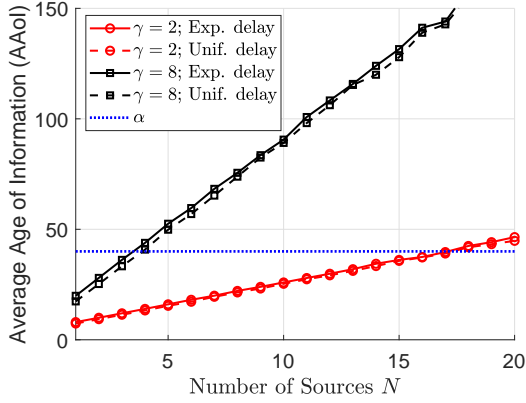}}
	\caption{Effect of number of sources on AAoI of an individual source.\vspace{-2ex}}
	\label{fig:Each_vsN}
\end{figure}

First, to analyze the effect of number of sources on the AAoI of an individual source, we consider a system with $N$ identical sources (for each source $\ell$, $\alpha_\ell=40$, $\mu_\ell=4$, and $\cD_\ell=\cD$), and simulate the system under policy $\pi_R$ for $N\in\{1,\cdots,20\}$, assuming $\cD$ to be $(i)$ an exponential  distribution with mean $\gamma\in\{2,8\}$, and $(ii)$ uniform distribution with mean $\gamma\in\{2,8\}$ (the rationale is to consider memoryless distribution, as well as a non-memoryless distribution $\cD$). Then, we plot the AAoI of source $1$ for different choices of $N$ and $\cD$ in Figure \ref{fig:Each_vsN}.

As shown in Figure \ref{fig:Each_vsN}, as $N$ increases, AAoI of source 1 increases linearly, with slope proportional to $\gamma$. This is because when sources are identical, the probability \eqref{eq:pick-prob} of picking a source for transmission is $p_\ell=1/N$, which implies the expected time interval between two successive instants when source $\ell$ gets to transmit is proportional to $N\gamma$ (since expected transmission delay for every transmitted packet is $\gamma$).    

\begin{figure} 
	\centerline{\includegraphics[width=0.85\linewidth]{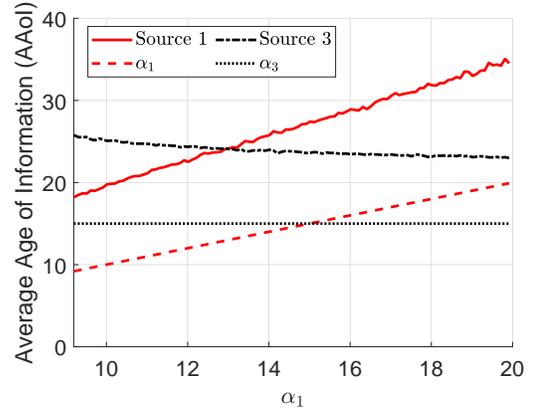}}
	\caption{Effect of parameter $\valpha$ on AAoI of sources.\vspace{-2ex}}
	\label{fig:Each_vsAlpha}
\end{figure}

Further, to analyze the effect of the recommended AAoI $\alpha_\ell$ of source $\ell$, we consider a system with $N=5$ sources, with mean packet inter-generation time $[\mu_1,\cdots,\mu_5]=[2,4,4,8,10]$, mean transmission delay $[\gamma_1,\cdots,\gamma_5]=[3,3,6,2,4]$, and $\valpha=[\alpha_1,10,15,20,20]$. Then, we simulate the system under policy $\pi_R$ for $\alpha_1\in[9.2,20]$, and plot the corresponding AAoI values for two of the sources 
in Figure \ref{fig:Each_vsAlpha}. 
\begin{remark}
The choice of parameters $\mu_\ell$, $\gamma_\ell$ and $\alpha_\ell$ ($\forall \ell$) are arbitrary, to avoid symmetry between sources. Further, we only consider $\alpha_1\ge 9.2$, because for the considered choice of other parameters, $\valpha\in\cC$, i.e., the conditions in Lemma \ref{lemma:NC} are satisfied, only when $\alpha_1\ge 9.2$
\end{remark}

Note that with increase in $\alpha_1$, $\av{T}_1^{\max}$ \eqref{eq:Tmax} increases, whereas $\av{T}_\ell^{\max}$ (for $\ell\ne 1$) decreases. Hence, $p_1$ \eqref{eq:pick-prob} (proportional to $1/\av{T}_1^{\max}$ decreases, while $p_\ell$ (for $\ell\ne 1$; inversely proportional to $1/\av{T}_1^{\max}$) increases with increase in $\alpha_1$. This is also illustrated in Figure \ref{fig:Each_vsAlpha}, where with increase in $\alpha_1$, AAoI of source $1$ increases, while AAoI of other source (source $3$) decreases. In addition, 
Figure \ref{fig:Each_vsAlpha} shows that 
when $\alpha_1\ge 9.2$ (i.e., when $\valpha\in\cC$), AAoI  of source $\ell\in\{1,3\}$ is less than $3\alpha_\ell$, which is expected because of Theorem \ref{thm:CR-piR}.

\begin{figure} 
	\centerline{\includegraphics[width=0.85\linewidth]{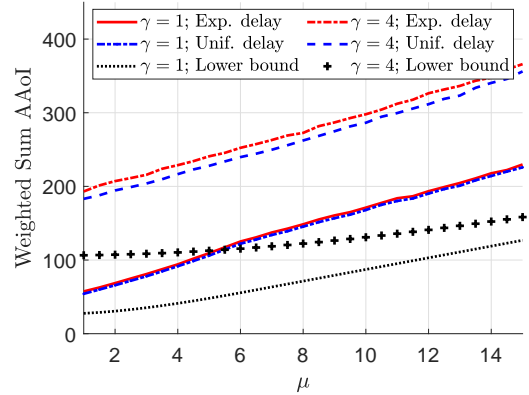}}
	\caption{Weighted sum AAoI as a function of mean packet inter-generation time.\vspace{-2ex}}
	\label{fig:Sum_vsArrival}
\end{figure}

Next, to verify Theorem \ref{thm:Sum-CR-piR}, we again consider a system with $N=5$ sources, and parameters $[\mu_1,\cdots,\mu_5]=\mu\cdot[2,4,4,8,10]$, mean transmission delay $[\gamma_1,\cdots,\gamma_5]=\gamma\cdot[3,3,6,2,4]$, and the relative weight vector $\bw=[0.8,0.8,0.2,0.2,0.4]$. Then, we simulate the system under policy $\pi_R$ for different values of $\mu$, $\gamma$ and family of transmission delay distribution $\cD_\ell$ (family of distribution is same for each source $\ell$), and plot the output in Figure \ref{fig:Sum_vsArrival}.
\begin{remark}
For simulating $\pi_R$ when the objective is to minimize the weighted Sum AAoI, we compute $\av{T}_\ell^o$ ($\forall \ell\in\{1,\cdots,N\}$) by solving \eqref{eq:lb-alpha-star-2} (using \emph{CVX} toolbox in Matlab), and use it to obtain $p_\ell^o$ by substituting $\av{T}_{\ell}^o$'s in \eqref{eq:p-ell-o}.
\end{remark}

It is evident from Figure \ref{fig:Sum_vsArrival} that the weighted sum AAoI for policy $\pi_R$ is less than $3$ times the theoretical lower bound $\sum_{\ell=1}^N w_\ell\alpha_\ell^o$ computed by solving the optimization problem \eqref{eq:lb-alpha-star-2}. Also, it can be noted that the mean transmission delay has significant impact on the weighted sum AAoI of $\pi_R$, whereas the family of transmission delay distribution $\cD_\ell$ (i.e., exponential or uniform) has comparatively negligible effect. Note that for any given value of the mean transmission delay for the sources, the lower bound \eqref{eq:lb-alpha-star-2} is independent of the family of 
distribution of $\cD_\ell$. 

Finally, we consider a system with single source ($N=1$) that can generate fresh packets at any time (i.e., the mean packet inter-generation time $\mu\to 0^+$), and each packet transmitted by the source suffers random transmission delay according to some general distribution $\cD$, with mean $\gamma$. 
Also, when $N=1$, since a policy does not need to choose among multiple sources, we consider a simplified version of $\pi_R$, where the source generates and transmits a packet whenever the channel is free, and the previously transmitted update is at least $\av{T}^o=\gamma$ \eqref{eq:lb-alpha-star-2} time units old. 
\begin{remark} \label{remark:piR-for-UpdateWait}
$\av{T}^o=\gamma$ is the inter-generation time of successive transmitted packets in the lower bound $\alpha^o \eqref{eq:alpha-o}$ (on AAoI of the source), when $N=1$ and $\mu\to 0^+$.
\end{remark}

\begin{figure} 
	\centerline{\includegraphics[width=0.85\linewidth]{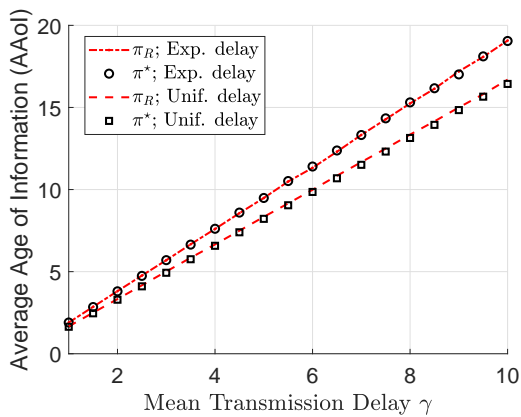}}
	\caption{$\pi_R$ versus $\pi^\star$ in a single source generate-at-will setup.\vspace{-2ex}}
	\label{fig:UpdateWait}
\end{figure}

We simulate the system under $\pi_R$, and an optimal policy $\pi^\star$ proposed in \cite{sun2017update}, for different choices of $\cD$ and $\gamma$, and plot the AAoI for the policies in Figure \ref{fig:UpdateWait}. Interestingly, for exponential and uniform distribution $\cD$, we find that the difference between the AAoI of the source under $\pi_R$ and $\pi^\star$ is negligible. 
This is despite the fact that $\pi_R$ is independent of the family of distribution that $\cD$ belongs to, and only depends on $\gamma$. Whereas, $\pi^\star$ is a threshold-based policy, where the threshold needs to be computed for each distribution $\cD$, which might be difficult for certain family of distributions. 


\section{Conclusion}
In this paper, we considered M/G/1 queuing model with multiple sources, where the objective is to minimize the expected average age of information (AAoI) of each source. 
We proposed an online randomized policy for prioritizing sources (and their packets), and showed that if there exists any online policy that can minimize expected AAoI of the sources below $\valpha$, 
then the proposed policy can guarantee expected AAoI of the sources, less than $3\valpha$. We further showed that in the setting where $\valpha$ is not known, and the objective is to minimize the weighted sum expected AAoI (WSAAoI) of the sources, the proposed policy guarantees WSAAoI that is at most $3$ times the minimum possible value (under any online policy). Using numerical simulations, we also showed that in special cases of the problem, where an optimal online policy is known, the proposed randomized policy might still be preferable due to ease of implementation, and near-optimal performance.

\bibliographystyle{IEEEtran}
\bibliography{reflist.bib}

\appendices
\section{Existence of policy $\pi\in\Pi$ for which the expected AAoI of each source is finite} \label{appendix:non-empty-PiON}

Recall that $\forall \ell\in\{1,\cdots,N\}$, $\mu_\ell$ and $\gamma_\ell$ are finite. Therefore, packet inter-generation times and transmission delays are finite with probability 1, for each source $\ell$. Hence, when $N$ is finite, for a round-robin policy, that picks a source, waits until a fresh packet is generated at the picked source, then transmits the generated packets, and then picks another source in cyclic order when channel becomes free, will have finite expected AAoI for each source.

\section{Proof of Lemma \ref{lemma:lower-bound} and Lemma \ref{lemma:ub-piR}} \label{appendix:proof-lemma-lb-ub}
Let $\ell_1^\pi,\ell_2^\pi,\ell_3^\pi,\cdots$ denote the sequence of packets of source $\ell$ that get transmitted under policy $\pi$. Also, let $g_{\ell i}^\pi$,
$s_{\ell i}^\pi$ and $r_{\ell i}^\pi$ respectively denote the generation time of packet $\ell_i^\pi$, time when transmission of packet $\ell_i^\pi$ begins, and the time when transmission of packet $\ell_i^\pi$ completes. Now, define $T_{\ell i}^\pi=g_{\ell i}^\pi-g_{\ell (i-1)}^\pi$, $W_{\ell i}^\pi=s_{\ell i}^\pi-g_{\ell i}^\pi$, and $d_{\ell i}=r_{\ell i}^\pi-s_{\ell i}^\pi$. Note that $T_{\ell i}^\pi$ is the inter-generation time of packets that are transmitted by policy $\pi$, and $Z_{\ell i}^\pi=W_{\ell i}^\pi+d_{\ell i}=r_{\ell i}^\pi-g_{\ell i}^\pi$ is the age of packet $\ell_i^\pi$ at the instant it is received at the monitor.  
\begin{remark} \label{remark:d-independent-pi}
Note that $d_{\ell i}$ is a random variable that denotes the transmission delay of packet $\ell_i^\pi$. By definition, $d_{\ell i}\sim\cD_\ell$ is independent of policy $\pi$.
\end{remark}
\begin{figure}
	\begin{center}
		\begin{tikzpicture}[thick,scale=0.8, every node/.style={scale=1}]
		\draw[->] (-0.25,0) to (7.8,0) node[below]{time ($t$)};
		\draw[->] (0.35,-0.1) node[below]{$0$} to (0.35,2.9) node[below left]{$\Delta_{\ell}(t)$};
		\draw (0.35,0) to (2.7,2.35) to (2.7,0.7) to (4.8,2.8) to (4.8,0.8) to (6.4,2.4) to (6.4,0.65) to (6.75,1); 
		
		
		\draw[loosely dotted] (7.0,1) to (7.7,1); 
		
		\draw (2,-0.1) node[below]{$g_{\ell 1}^\pi$} to (2,0.1);
		\draw (4,-0.1) node[below]{$g_{\ell 2}^\pi$} to (4,0.1);
		\draw (5.75,-0.1) node[below]{$g_{\ell 3}^\pi$} to (5.75,0.1); 
		
		\draw (2.7,-0.1) node[below]{$r_{\ell 1}^\pi$} to (2.7,0.1);
		\draw (4.8,-0.1) node[below]{$r_{\ell 2}^\pi$} to (4.8,0.1);
		\draw (6.4,-0.1) node[below]{$r_{\ell 3}^\pi$} to (6.4,0.1); 
		
		\draw[dashed] (2,0.1) to (2,2.7);
		\draw[dashed] (2.7,0.35) to (2.7,2.7);
		\draw[dashed] (4,0.1) to (4,2.7);
		\draw[dashed] (5.75,0.1) to (5.75,2.7);
		
		\draw[dashed] (2,0) to (2.7,0.7) to (2.7,0.1);
		\draw[dashed] (4,0) to (4.8,0.8) to (4.8,0.1);
		\draw[dashed] (5.75,0) to (6.4,0.65) to (6.4,0.1);
		\draw[dashed] (6.4,2.4) to (6.4,2.7);
		
		\draw[|<->] (0.35,-1) -- (2,-1) node[rectangle,inner sep=-1pt,midway,fill=white]{$T_{\ell 1}$}; 
		\draw[|<->] (2,-1) -- (4,-1) node[rectangle,inner sep=-1pt,midway,fill=white]{$T_{\ell 2}$};
		\draw[|<->|<] (4,-1) -- (5.8,-1) node[rectangle,inner sep=-1pt,midway,fill=white]{$T_{\ell 3}$};
		
		\draw[|<->|] (2,3) -- (2.7,3) node[rectangle,inner sep=-1pt,midway,fill=white]{$Z_{\ell 1}$}; 
		\draw[|<->|] (4,3) -- (4.8,3) node[rectangle,inner sep=-1pt,midway,fill=white]{$Z_{\ell 2}$};
		\draw[|<->|] (5.75,3) -- (6.4,3) node[rectangle,inner sep=-1pt,midway,fill=white]{$Z_{\ell 3}$};
		
		\draw[loosely dotted] (6,-1) to (6.8,-1);
		
		\end{tikzpicture}
		\caption{Sample AoI plot of source $\ell$ in a multi-source setting.  
		Here, a packet $\ell_i^\pi$, generated at time $g_{\ell i}^\pi$, is received at the monitor at time $r_{\ell i}^\pi$.\vspace{-2ex}
		}
		\label{fig:notations} 
	\end{center}
\end{figure}
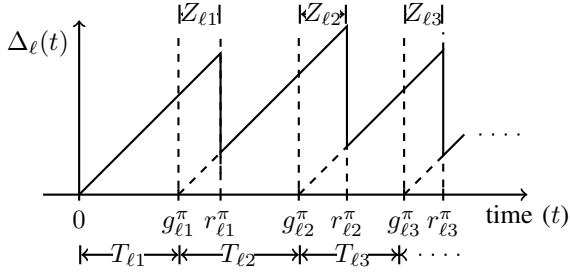

Figure \ref{fig:notations} shows a sample AoI plot for source $\ell$, labelled with the quantities defined above. As evident from Figure \ref{fig:notations} (and shown in detail in \cite{kaul2012status}), AAoI \eqref{eq:aaoi} of source $\ell$ can be expressed in terms of the defined quantities as follows
\begin{align} \label{eq:AAoI-pi}
    \lim_{t\to\infty}\aaoi_\ell^\pi(t)\stackrel{(a)}{=}\lim_{t\to\infty}\frac{\sum_{i=1}^{R_\ell^\pi(t)}((T_{\ell i}^\pi)^2/2+T_{\ell i}^\pi Z_{\ell i}^\pi)}{t}, 
\end{align}
where $R_{\ell}^\pi(t)$ denotes the number of packets transmitted by source $\ell$ under policy $\pi$, until time $t$.  
Also, $t=\sum_{i=1}^{R_{\ell}^\pi(t)}T_{\ell i}^\pi$ (shown in \cite{saurav2021minimizing}). Therefore,  
\begin{align} \label{eq:Tav}
    \lim_{t\to\infty}\frac{\sum_{i=1}^{R_{\ell}^\pi(t)}T_{\ell i}^\pi}{R_\ell^\pi(t)}=\lim_{t\to\infty}\frac{t}{R_\ell^\pi(t)}
    =\av{T}_\ell^\pi,
\end{align}
where $\av{T}_\ell^\pi$ denotes the average inter-generation time of packets transmitted by policy $\pi$.

From \eqref{eq:Tav}, it follows that when $t\to\infty$, $\sum_{\ell=1}^{R_\ell^\pi(t)}T_{\ell i}^\pi=R_\ell^\pi(t)\av{T}_\ell^\pi$. 
Therefore, 
\begin{align} \label{eq:delta-sum-0}
    \sum_{i=1}^{R_\ell^\pi(t)}T_{\ell i}^\pi-R_\ell^\pi(t)\av{T}_\ell^\pi=\sum_{i=1}^{R_\ell^\pi(t)}(T_{\ell i}^\pi-\av{T}_\ell^\pi)=\sum_{i=1}^{R_\ell^\pi(t)}\delta_{\ell i}^\pi=0
\end{align}
where $\delta_{\ell i}^\pi=T_{\ell i}^\pi-\av{T}_\ell^\pi$.
This also implies that when $t\to\infty$,
\begin{align} \label{eq:sum-T}
    \sum_{i=1}^{R_\ell^\pi(t)}T_{\ell i}^\pi=\sum_{i=1}^{R_\ell^\pi(t)}\delta_{\ell i}^\pi+\sum_{i=1}^{R_\ell^\pi(t)}\av{T}_{\ell}^\pi=R_\ell^\pi(t)\cdot\av{T}_{\ell}^\pi.
\end{align}

Further, squaring both sides of $T_{\ell i}^\pi=\delta_{\ell i}^\pi+\av{T}_{\ell}^\pi$, we get $(T_{\ell i}^\pi)^2=(\delta_{\ell i}^\pi)^2+(\av{T}_{\ell}^\pi)^2+2\av{T}_\ell^\pi\delta_{\ell i}^\pi$. Hence, 
\begin{align} \label{eq:sum-T2}
    \sum_{i=1}^{R_\ell^\pi(t)}(T_{\ell i}^\pi)^2&=\sum_{i=1}^{R_\ell^\pi(t)}(\delta_{\ell i}^\pi)^2+\sum_{i=1}^{R_\ell^\pi(t)}(\av{T}_{\ell}^\pi)^2+2\av{T}_{\ell}^\pi\sum_{i=1}^{R_\ell^\pi(t)}\delta_{\ell i}^\pi, \nonumber \\
    &\stackrel{(a)}{=}\sum_{i=1}^{R_\ell^\pi(t)}(\delta_{\ell i}^\pi)^2+\sum_{i=1}^{R_\ell^\pi(t)}(\av{T}_{\ell}^\pi)^2, \nonumber \\
    &=\sum_{i=1}^{R_\ell^\pi(t)}(\delta_{\ell i}^\pi)^2+R_\ell^\pi(t)\cdot(\av{T}_{\ell}^\pi)^2,
\end{align}
where we get (a), because $\sum_{i=1}^{R_\ell^\pi(t)}\delta_{\ell i}^\pi=0$ (from \eqref{eq:delta-sum-0}).

Substituting \eqref{eq:sum-T} and \eqref{eq:sum-T2} in \eqref{eq:AAoI-pi}, and using the relation $t=\sum_{i=1}^{R_\ell^\pi(t)}T_{\ell i}^\pi=R_\ell^\pi(t)\cdot \av{T}_\ell^\pi$ (for $t\to\infty$), 
we get $\lim_{t\to\infty}\aaoi_\ell^\pi(t)$
\begin{align} \label{eq:AAoI-beta-phi}
    &=\lim_{t\to\infty}\left(\frac{\sum_{i=1}^{R_\ell^\pi(t)}(\delta_{\ell i}^\pi)^2}{2R_\ell^\pi(t)\cdot\av{T}_\ell^\pi}+\frac{\av{T}_\ell^\pi}{2}+\frac{\sum_{i=1}^{R_\ell^\pi(t)}T_{\ell i}^\pi Z_{\ell i}^\pi}{\sum_{i=1}^{R_\ell^\pi(t)}T_{\ell i}^\pi}\right),\nonumber \\
    &=\frac{\beta_\ell^\pi}{2\av{T}_\ell^\pi}+\frac{\av{T}_\ell^\pi}{2}+\phi_\ell^\pi,
\end{align}
where 
\begin{align}
    \label{eq:beta}
    \beta_\ell^\pi&=\lim_{t\to\infty}\frac{\sum_{i=1}^{R_\ell^\pi(t)}(\delta_{\ell i}^\pi)^2}{R_\ell^\pi(t)}, \text{ and } \\
    \label{eq:phi}
    \phi_\ell^\pi&=\lim_{t\to\infty}\frac{\sum_{i=1}^{R_\ell^\pi(t)}T_{\ell i}^\pi Z_{\ell i}^\pi}{\sum_{i=1}^{R_\ell^\pi(t)}T_{\ell i}^\pi}.
\end{align}

Taking expectation on both sides of \eqref{eq:AAoI-beta-phi}, 
we get
\begin{align} \label{eq:exp-aaoi-pre}
    \bbE_\pi\left[\lim_{t\to\infty}\aaoi_\ell^\pi(t)\right]
    =\bbE_\pi\left[\frac{\beta_\ell^\pi}{2\av{T}_\ell^\pi}+\frac{\av{T}_\ell^\pi}{2}\right]+\bbE_\pi[\phi_\ell^\pi].
\end{align} 

\begin{proposition} \label{prop:avDel-expDel}
For any policy $\pi\in\Pi$, $(i)$ $\av{T}_\ell^\pi$ is finite, and $(ii)$ as $t\to\infty$, $R_\ell^\pi(t)\to\infty$ as well. 
\end{proposition}
\begin{IEEEproof}
$(i)$ As discussed in Remark \ref{remark:PiON}, for any policy $\pi\in\Pi$, AAoI of each source is finite. Hence, from \eqref{eq:AAoI-beta-phi} we get that for any $\pi\in\Pi$, $\av{T}_\ell^\pi$ is finite for each source $\ell$.
$(ii)$ From \eqref{eq:Tav}, we know that when $t$ is large, we can write it as $t=R_\ell^\pi(t)\cdot \av{T}_\ell^\pi$. Also, in $(i)$, we showed that $\av{T}_\ell^\pi$ is finite. Hence, when $t\to\infty$, then $R_\ell^\pi(t)\to\infty$ as well. 
\end{IEEEproof} 

As shown in Proposition \ref{prop:avDel-expDel}, as $t\to\infty$, $R_\ell^\pi(t)\to\infty$, $\forall \pi\in\Pi$. Hence, 
\begin{align} \label{eq:exp-phi-pre}
    \bbE_\pi[\phi_{\ell i}^\pi]
    &=\bbE_\pi\left[\lim_{t\to\infty}\frac{\sum_{i=1}^{R_\ell^\pi(t)}T_{\ell i}^\pi Z_{\ell i}^\pi}{\sum_{i=1}^{R_\ell^\pi(t)}T_{\ell i}^\pi}\right], \nonumber \\
    &=\bbE_\pi\left[\frac{\sum_{i=1}^{\infty}T_{\ell i}^\pi Z_{\ell i}^\pi}{\sum_{i=1}^{\infty}T_{\ell i}^\pi}\right], \nonumber \\
    &=\bbE_\pi\left[\sum_{i=1}^{\infty}\psi_{\ell i}^\pi Z_{\ell i}^\pi\right], 
\end{align}
where 
\begin{align} \label{eq:psi}
    \psi_{\ell i}^\pi=\frac{T_{\ell i}^\pi}{\sum_{j=1}^\infty T_{\ell j}^\pi}. 
\end{align}
Thus, for each source $\ell$, and any policy $\pi\in\Pi$, $\psi_{\ell i}^\pi\in[0,1]$ ($\forall i\ge 1$), and $\sum_{i=1}^\infty \psi_{\ell i}^\pi=1$. 

Using Tonelli's Theorem \cite{patrick1995probability} to exchange infinite sum and expectation on the R.H.S. of \eqref{eq:exp-phi-pre} (by definition, each term in \eqref{eq:exp-phi-pre} is non-negative and measurable), we get 
\begin{align} \label{eq:exp-phi}
    \bbE_\pi[\phi_{\ell i}^\pi]
    &=\sum_{i=1}^{\infty}\bbE_\pi\left[\psi_{\ell i}^\pi Z_{\ell i}^\pi\right]. 
\end{align} 

Substituting \eqref{eq:exp-phi} in \eqref{eq:exp-aaoi-pre}, we get
\begin{align} \label{eq:exp-aaoi}
    \bbE_\pi\left[\lim_{t\to\infty}\Delta_\ell^\pi(t)\right]
    &=\bbE_\pi\left[\frac{\beta_\ell^\pi}{2\av{T}_\ell^\pi}+\frac{\av{T}_\ell^\pi}{2}\right]+\sum_{i=1}^{\infty}\bbE_\pi\left[\psi_{\ell i}^\pi Z_{\ell i}^\pi\right]. 
\end{align}

\subsection{Proof of Lemma \ref{lemma:lower-bound}} 

Recall that $Z_{\ell i}^\pi=W_{\ell i}^\pi+d_{\ell i}$, where $W_{\ell i}^\pi=s_{\ell i}^\pi-g_{\ell i}^\pi\ge 0$, and $d_{\ell i}$'s are independent and identically distributed according to distribution $\cD_\ell$. Also, $\psi_{\ell i}^\pi$'s are non-negative (by definition).
Therefore, $\psi_{\ell i}^\pi Z_{\ell i}^\pi\ge \psi_{\ell i}^\pi d_{\ell i}$, which implies $\bbE_\pi[\psi_{\ell i}^\pi Z_{\ell i}^\pi]\ge \bbE[\psi_{\ell i}^\pi d_{\ell i}]$. 

Further, note that when a policy $\pi$ initiates the transmission of packet $\ell_i^\pi$ at time $s_{\ell i}^\pi$, $T_{\ell i}^\pi$ gets fixed, whereas $d_{\ell i}\sim\cD_\ell$ is realized when the transmission of packet $\ell_i^\pi$ completes at time $r_{\ell i}^\pi$, independent of everything that happened until time $s_{\ell i}^\pi$. Hence, $T_{\ell i}^\pi$ and $d_{\ell i}$ are mutually independent. Also, $\sum_{i=1}^{\infty}T_{\ell i}^\pi$ is equal to the time horizon (from \eqref{eq:Tav}), independent of everything else.  
Therefore, $\psi_{\ell i}^\pi$ \eqref{eq:psi} and $d_{\ell i}$ are mutually independent. 
Hence, $\bbE_\pi[\psi_{\ell i}^\pi d_{\ell i}]=\bbE_\pi[\psi_{\ell i}^\pi]\bbE_\pi[d_{\ell i}]=\bbE_\pi[\psi_{\ell i}^\pi]$ (mean of $d_\ell\sim\cD_\ell$ is $\gamma_\ell$). Hence, 
\begin{align} \label{eq:psi-to-gamma}
    \sum_{i=1}^\infty\bbE_\pi\left[\psi_{\ell i}^\pi Z_{\ell i}^\pi\right]&\ge \sum_{i=1}^\infty\bbE_\pi\left[\psi_{\ell i}^\pi d_{\ell i}^\pi\right], \nonumber \\
    &= \sum_{i=1}^\infty\bbE_\pi[\psi_{\ell i}^\pi]\gamma_\ell, \nonumber \\
    &\stackrel{(a)}{=} \bbE_\pi\left[\sum_{i=1}^\infty\psi_{\ell i}^\pi\right]\gamma_\ell, \nonumber \\
    &\stackrel{(b)}{=}\gamma_\ell,
\end{align}
where we get $(a)$ using Tonelli's Theorem \cite{patrick1995probability}, and $(b)$ follows because $\sum_{i=1}^{\infty}\psi_{\ell i}^\pi=1$ (by definition; \eqref{eq:psi}).

Also, $\beta_\ell^\pi\ge \mu_\ell^2/2$, as shown in the following proposition. 
\begin{proposition} \label{prop:lb-beta}
   For any $\pi\in\Pi$, $\beta_\ell^\pi\ge \mu_\ell^2/2$, for each source $\ell$. 
\end{proposition}
\begin{IEEEproof}
See Appendix \ref{appendix:prop-lb-beta}.
\end{IEEEproof}

 From \eqref{eq:exp-aaoi}, \eqref{eq:psi-to-gamma} and Proposition \ref{prop:lb-beta}, 
 we get 
 \begin{align} 
		\bbE_\pi\left[\lim_{t\to\infty}\aaoi_\ell(t)\right]&\ge \frac{1}{2}\left(\bbE_\pi\left[\frac{\mu_\ell^2/2}{\av{T}_\ell^\pi}+\av{T}_\ell^\pi\right]+2\gamma_\ell\right), \nonumber \\
		&= \frac{1}{2}\left(\bbE_\pi\left[\frac{\mu_\ell^2/2}{\av{T}_\ell^\pi}\right]+\bbE_\pi\left[\av{T}_\ell^\pi\right]+2\gamma_\ell\right), \nonumber \\
		&\stackrel{(a)}{\ge} \frac{1}{2}\left(\frac{\mu_\ell^2/2}{\bbE_\pi[\av{T}_\ell^\pi]}+\bbE_\pi[\av{T}_\ell^\pi]+2\gamma_\ell\right),
 \end{align}
where $(a)$ follows from Jensen's inequality. 

Further, note that since at most one source can transmit at a time, and each transmission by source $\ell$ keeps the channel busy for $d_\ell\sim\cD_\ell$ time units, we get $\lim_{t\to\infty}\sum_{\ell=1}^N\sum_{i=1}^{R_\ell^\pi(t)}d_{\ell i}\le t$. Hence, 
\begin{align} \label{eq:throughput-constraint-proof-1}
    1&\ge\lim_{t\to\infty}\sum_{\ell=1}^N\left(\frac{\sum_{i=1}^{R_\ell^\pi(t)}d_{\ell i}}{R_\ell^\pi(t)}\cdot\frac{R_{\ell}^\pi(t)}{t}\right), \nonumber \\ 
    &\stackrel{(a)}{=}\sum_{\ell=1}^N\left(\lim_{t\to\infty}\frac{\sum_{i=1}^{R_\ell^\pi(t)}d_{\ell i}}{R_\ell^\pi(t)}\cdot\lim_{t\to\infty}\frac{R_{\ell}^\pi(t)}{t}\right), \nonumber \\
    &\stackrel{(b)}{=}\sum_{\ell=1}^N\frac{\gamma_\ell}{\av{T}_\ell^\pi}, 
\end{align} 
where we get $(a)$, because limit of products is equal to product of limits (when limits exists, as in the above case), and $(b)$ follows because $\lim_{t\to\infty} R_\ell^\pi(t)/t=1/\av{T}_\ell^\pi$ (from \eqref{eq:Tav}), and  $\lim_{t\to\infty}\sum_{i=1}^{R_\ell^\pi(t)}d_{\ell i}/R_\ell^\pi(t)=\gamma_\ell$ with probability 1 (using strong law of large number; $\lim_{t\to\infty}R_\ell^\pi(t)\to\infty$ from Proposition \ref{prop:avDel-expDel}). 
Finally, taking expectation on both sides of \eqref{eq:throughput-constraint-proof-1}, and using Jensen's inequality, we get
\begin{align}
    1\ge \sum_{\ell=1}^N\frac{\gamma_\ell}{\bbE_\pi[\av{T}_\ell^\pi]}.
\end{align}

\subsection{Proof of Lemma \ref{lemma:ub-piR}} 

For each source $\ell$, consider the sequence of packets marked by the following threshold policy $\pi_{Th}$: Mark the first packet generated at source $\ell$, and subsequently, whenever a packet is generated at the source, mark the packet if the time elapsed since the generation time of the previously marked packet is at least $\av{T}_{\ell}^{\max}-\mu_\ell$ (otherwise, discard the packet). Let this sequence of marked packets be denoted by $\cM_\ell$. 

Now, consider a modified randomized policy $\pi_M$, which is identical to the randomized policy $\pi_R$ (Algorithm \ref{algo:threshold-policy}), except that whenever a source $\ell$ is picked to transmit its packet, instead of transmitting the latest generated packet, $\pi_M$ transmits the latest generated packet from $\cM_\ell$ (and idles for $d_\ell$ time units, if there is no packet from $\cM_\ell$ to transmit). As shown in \cite{saurav2021minimizing,kaul2012status}, the AAoI of source $\ell$ under policy $\pi_R$ is at most equal to AAoI of policy $\pi_M$, because transmitting latest generated packet causes at least as much reduction in AoI, as transmitting the latest packet from $\cM_\ell$ (and everything else remains the same for the two policies). Hence, to prove Lemma \ref{lemma:ub-piR} (upper bound \eqref{eq:ub-piR} on the expected AAoI of source $\ell$ under policy $\pi_R$), it is sufficient to show that the upper bound \eqref{eq:ub-piR} is applicable for the expected AAoI of the sources under policy $\pi_M$. 

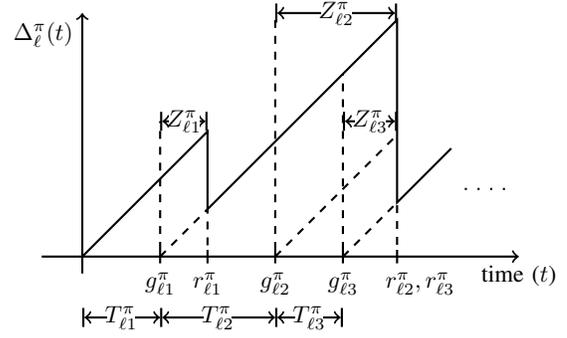
\begin{figure}
	\begin{center}
		\begin{tikzpicture}[thick,scale=0.9, every node/.style={scale=0.9}]
		\draw[->] (-0.25,0) to (6.8,0) node[below]{time ($t$)};
		\draw[->] (0.35,-0.25) to (0.35,3.6) node[below left]{$\Delta_{\ell}^{\pi}(t)$};
		\draw (0.35,0) to (2.2,1.85) to (2.2,0.7) to (5,3.5) to (5,0.8) to (5.75,1.55) to (5.8,1.6) ; 
		
		
		\draw[loosely dotted] (6,1) to (6.7,1); 

		\draw (1.5,-0.1) node[below]{$g_{\ell 1}^\pi$} to (1.5,0.1);
		\draw (3.2,-0.1) node[below]{$g_{\ell 2}^\pi$} to (3.2,0.1);
		\draw (4.2,-0.1) node[below]{$g_{\ell 3}^\pi$} to (4.2,0.1); 
		
		\draw (2.2,-0.1) node[below]{$r_{\ell 1}^\pi$} to (2.2,0.1);
		\draw (5,-0.1) to (5,0.1);
		\draw (4.7,-0.1) node[below right]{$r_{\ell 2}^\pi,r_{\ell 3}^\pi$};
		
		\draw[dashed] (1.5,0.1) to (1.5,2);
		\draw[dashed] (3.2,0.1) to (3.2,3.6);
		\draw[dashed] (4.2,0.1) to (4.2,2.7);
		
		\draw[dashed] (1.5,0) to (2.2,0.7) to (2.2,0.1);
		\draw[dashed] (3.2,0) to (5,1.8); 
		\draw[dashed] (4.2,0) to (5,0.8) to (5.,0.1);
		
        \draw[|<->] (0.35,-0.9) -- (1.5,-0.9) node[rectangle,inner sep=-1pt,midway,fill=white]{$T_{\ell 1}^\pi$}; 
        \draw[|<->] (1.5,-0.9) -- (3.2,-0.9) node[rectangle,inner sep=-1pt,midway,fill=white]{$T_{\ell 2}^\pi$};
        \draw[|<->|] (3.2,-0.9) -- (4.2,-0.9) node[rectangle,inner sep=-1pt,midway,fill=white]{$T_{\ell 3}^\pi$};
		
        \draw[|<->|] (1.5,2) -- (2.2,2) node[rectangle,inner sep=-1pt,midway,fill=white]{$Z_{\ell 1}^\pi$}; 
        \draw[|<->|] (3.2,3.6) -- (5,3.6) node[rectangle,inner sep=-1pt,midway,fill=white]{$Z_{\ell 2}^\pi$};
        \draw[|<->|] (4.2,2) -- (5,2) node[rectangle,inner sep=-1pt,midway,fill=white]{$Z_{\ell 3}^\pi$};
		
		
		\end{tikzpicture}
		\caption{Sample AoI plot of source $\ell$ under policy $\pi=\pi_M$. The packet generated at time $g_{\ell 2}^\pi$ can be considered to have been transmitted simultaneously with the packet generated at time $g_{\ell 3}^\pi$. Also, the variables $T_{\ell i}^\pi$ and $Z_{\ell i}^\pi$ can be defined accordingly. \vspace{-4ex} 
		} 
		\label{fig:age-threshold-randomized-policy} 
	\end{center}
\end{figure}

Note that from AoI perspective, transmitting a packet $\ell_i^\pi$ generated at time $g_{\ell i}^\pi$ is equivalent to simultaneously transmitting packet $\ell_i^\pi$, and the packets that were generated before $g_{\ell i}^\pi$ (all packets get received at the monitor simultaneously at time $r_{\ell i}^\pi$). This is because as shown in Figure \ref{fig:age-threshold-randomized-policy}, the reduction in AoI at time $r_{\ell i}^\pi$ is equal in both the cases. Hence, we interpret $\pi_M$ as follows: whenever $\pi_R$ picks a source $\ell$ to transmit, $\pi_M$ simultaneously transmits all the available packets from the set $\cM_\ell$, that were generated after the previous instant when source $\ell$ got to transmit. 
\begin{remark} \label{remark:pith-piR}
Note that $\pi_{Th}$ and $\pi_R$ are mutually independent policies. $\pi_M$ uses these two policies as subroutines, to decide which of the generated packets are transmitted, and at what time instant. Nonetheless, under the above-mentioned interpretation, under policy $\pi_M$, the packets that are transmitted, and the time-instants when the packets get transmitted, are mutually independent.
\end{remark}

Thus, for policy $\pi_M$, we consider the generation time of transmitted packets, i.e., $g_{\ell 1}^M, g_{\ell 2}^M, \cdots$ to be same as the sequence of generation time of packets in $\cM_\ell$. Accordingly, for the packets, we define $s_{\ell i}^M$ and $r_{\ell i}^M$ respectively, as the time instants when the transmission of these packets start and complete (its just that for some of the successive packets, these time instants may be the same, as shown in Figure \ref{fig:age-threshold-randomized-policy} (and later in Figure \ref{fig:R-th-policy}).  Also, $T_{\ell i}^M=g_{\ell i}^M-g_{\ell (i-1)}^M$, and $Z_{\ell i}^M=W_{\ell i}^M+d_{\ell i}$, where $W_{\ell i}^M=s_{\ell i}^M-g_{\ell i}^M$ and $d_{\ell i}=r_{\ell i}^M-s_{\ell i}^M$.

\begin{proposition} \label{prop:Tav-beta-M}
For each source $\ell$, 
\begin{enumerate}
    \item $\av{T}_{\ell}^M=\bbE_M[T_{\ell i}^M]=\av{T}_{\ell}^{max}$, where $\av{T}_{\ell}^M$ \eqref{eq:Tav} is the average inter-generation time $T_{\ell i}^M$ of packets transmitted by $\pi_M$ (marked by $\pi_{Th}$). 
    \item $\beta_\ell^M=\mu_\ell^2$, where $\beta_\ell^M$ is the empirical variance \eqref{eq:beta} of $T_{\ell i}^M$ (for $i\ge1$).
    \item $\bbE_M[Z_{\ell i}^M]=\bbE_M[W_{\ell i}^M]+\bbE_M[d_{\ell i}]\le \av{T}_{\ell}^{\max}+\gamma_\ell$.
\end{enumerate}
\end{proposition}
\begin{IEEEproof}
See Appendix \ref{Appendix-proof-prop-Tav-beta-M}.
\end{IEEEproof}

From Proposition \ref{prop:Tav-beta-M}, it follows that the inter-generation time of successive packets that $\pi_M$ transmits is bounded with probability 1 (since $\av{T}_\ell^{\max}$ is bounded; Remark \ref{remark:Tmax-finite}). Also, the difference between the time when these packets are generated to when they are received at the monitor (i.e., $Z_{\ell i}^M$) are finite with probability 1 (because $\bbE_M[Z_{\ell i}^M]$ is finite). 
Therefore, the AAoI for each source under $\pi_M$ must be finite. 
Hence, $\pi_M\in\Pi$ (follows from the definition of $\Pi$; Remark \ref{remark:PiON}). 
Thus, using relation \eqref{eq:exp-aaoi} (which is true for any policy in $\Pi$), we get $\bbE_M\left[\lim_{t\to\infty}\aaoi_\ell^M(t)\right]$
\begin{align} \label{eq:exp-aaoi-R}
    &=\bbE_M\left[\frac{\beta_\ell^M}{2\av{T}_\ell^M}+\frac{\av{T}_\ell^M}{2}\right]+\sum_{i=1}^{\infty}\bbE_M\left[\psi_{\ell i}^M Z_{\ell i}^M\right], \nonumber \\
    &\stackrel{(a)}{=}\frac{\mu_\ell^2}{2\av{T}_\ell^{\max}}+\frac{\av{T}_\ell^{\max}}{2}+\sum_{i=1}^{\infty}\bbE_M\left[\psi_{\ell i}^M Z_{\ell i}^M\right],
\end{align}
where $(a)$ follows from Proposition \ref{prop:Tav-beta-M}. Using \eqref{eq:exp-aaoi-R} and Proposition \ref{prop:psi-M} (discussed below), we get Lemma \ref{lemma:ub-piR}.

\begin{proposition} \label{prop:psi-M}
Under policy $\pi_M$, for each source $\ell$,
\begin{align}
    \sum_{i=1}^{\infty}\bbE_M\left[\psi_{\ell i}^M Z_{\ell i}^M\right]=\av{T}_\ell^{\max}+\gamma_\ell.
\end{align}
\end{proposition}
\begin{IEEEproof}
From \eqref{eq:psi-to-gamma}, recall that $\psi_{\ell i}^M=T_{\ell i}^M/\sum_{j=1}^{\infty}T_{\ell j}^M$. 
Note that $T_{\ell i}^M$ gets fixed when $\pi_{Th}$ marks packet $\ell_i^M$ at its generation time $g_{\ell i}^M$, whereas $W_{\ell i}^M$ and $d_{\ell i}$ are realized after $g_{\ell i}^M$, depending on when $\pi_R$ picks source $\ell$ (which is independent of $\pi_{Th}$). Hence, $T_{\ell i}^M$ and $Z_{\ell i}^M=W_{\ell i}^M+d_{\ell i}$ are mutually independent. Therefore, $\psi_{\ell i}^M(t)$ and $Z_{\ell i}^M$ are mutually independent. Also, $\pi_M$ is a stationary policy, and hence $Z_{\ell i}^M$ are identically distributed $\forall i\ge1$. Thus, $\bbE_M[\psi_{\ell i}^M Z_{\ell i}^M]=\bbE_M[\psi_{\ell i}^M]\bbE_M[Z_{\ell i}^M]$. 
Hence,
\begin{align}
    \sum_{i=1}^\infty \bbE_M\left[\psi_{\ell i}^M Z_{\ell i}^M\right]
    &=\sum_{i=1}^{\infty}\bbE_M[\psi_{\ell i}^M]\bbE_M[Z_{\ell i}^M], \nonumber \\
    &\stackrel{(a)}{=}\bbE_M\left[\sum_{i=1}^{\infty}\psi_{\ell i}^M\right]\bbE_M[Z_{\ell i}^M], \nonumber \\
    &\stackrel{b}{=}\bbE_M[W_{\ell i}^M]+\bbE_M[d_{\ell i}], \nonumber \\ 
    &\stackrel{(c)}{\le} \av{T}_\ell^{\max}+\gamma_\ell, 
\end{align} 
where we get $(a)$ using Tonelli's Theorem \cite{patrick1995probability}, $(b)$ follows from the definition of $\psi_{\ell i}^\pi$ \eqref{eq:psi}, and we get $(c)$ using Proposition \ref{prop:Tav-beta-M}.
\end{IEEEproof}

\section{Proof of Proposition \ref{prop:lb-beta}} \label{appendix:prop-lb-beta}
\begin{figure} 
	\begin{center}
		\begin{tikzpicture}[thick,scale=0.8, every node/.style={scale=1}]
		\draw[->] (-0.25,0) to (7.3,0) node[below]{$t$};
		
		
		\draw (0.35,-0.1) node[below]{$g_{\ell (i-1)}^\pi$} to (0.35,0.1);
		\draw (2.7,-0.1) to (2.7,0.1);
		\draw (4,-0.1) to (4,0.1);
        \draw (5.3,-0.1) to (5.3,0.1); 
        \draw (6,-0.1) node[below]{$g_{\ell i}^\pi$} to (6,0.1);

		\draw[|<->|] (0.35,0.4) -- (4,0.4) node[rectangle,inner sep=-1pt,midway,fill=white]{$\av{T}_{\ell}^\pi$}; 
		\draw[|<->|] (0.35,1) -- (6,1) node[rectangle,inner sep=-1pt,midway,fill=white]{$T_{\ell i}^\pi$};
		\draw[<->|] (4,0.4) -- (6,0.4) node[rectangle,inner sep=-1pt,midway,fill=white]{$\delta_{\ell i}^\pi$};
		\draw[|<->|] (2.7,-0.4) -- (4,-0.4) node[rectangle,inner sep=-1pt,midway,fill=white]{$\epsilon_{\ell i}$}; 
		\draw[<->|] (4,-0.4) -- (5.3,-0.4) node[rectangle,inner sep=-1pt,midway,fill=white]{$\epsilon_{\ell i}$};
		
		
		
		\end{tikzpicture}
		\caption{If no packet is generated in interval $(g_{\ell (i-1)}^\pi+\av{T}_{\ell}^\pi-\epsilon_{\ell i},g_{\ell (i-1)}^\pi+\av{T}_{\ell}^\pi+\epsilon_{\ell i})$, then $(\delta_{\ell i}^\pi)^2=(T_{\ell i}^\pi-\av{T}_{\ell}^\pi)^2\ge\epsilon_{\ell i}^2$.\vspace{-2ex}}  
		\label{fig:cycle-var} 
	\end{center}
\end{figure}
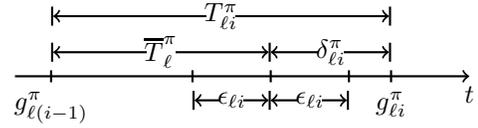
Consider the time interval $(g_{\ell (i-1)}^\pi,g_{\ell i}^\pi]$ of length $\av{T}_{\ell i}^\pi=g_{\ell i}^\pi-g_{\ell (i-1)}^\pi$. 
Since $g_{\ell i}^\pi$ is the generation time of a packet, if no packet is generated in interval $(g_{\ell (i-1)}^\pi+\av{T}_{\ell}^\pi-\epsilon_{\ell i},g_{\ell (i-1)}^\pi+\av{T}_{\ell}^\pi+\epsilon_{\ell i})$, then as shown in Figure \ref{fig:cycle-var}, we must have $(T_{\ell i}^\pi-\av{T}_{\ell}^\pi)^2=(\delta_{\ell i}^\pi)^2\ge\epsilon_{\ell i}^2$. Therefore, 
\begin{align} \label{eq:beta-lb-1}
\beta_\ell^\pi=\lim_{t\to\infty}\frac{\sum_{i=1}^{R_\ell^\pi(t)}(\delta_{\ell i}^\pi)^2}{R_\ell^\pi(t)}\ge\lim_{t\to\infty}\frac{\sum_{i=1}^{R_\ell^\pi(t)}\epsilon_{\ell i}^2}{R_\ell^\pi(t)},
\end{align}
where 
$\epsilon_{\ell i}$ is a random variable such that no packet is generated in interval $(g_{\ell (i-1)}^\pi+\av{T}_{\ell}^\pi-\epsilon_{\ell i},g_{\ell (i-1)}^\pi+\av{T}_{\ell}^\pi+\epsilon_{\ell i})$.  

Note that for any given sequence of packet inter-generation times, and policy $\pi\in\Pi$, $g_{\ell (i-1)}^\pi$ and $\av{T}_{\ell}^\pi$ are fixed, and $\epsilon_{\ell i}$ ($\forall i\ge1$) are independent and identically distributed. In particular, because inter-generation time of packets is exponentially distributed with mean $\mu_\ell$, number of packets generated in an interval of length $2\epsilon$ (where $\epsilon\ge0$) follows Poisson distribution with parameter $1/\mu_\ell$, and hence, the probability that $\epsilon_{\ell i}^2>\epsilon^2$ (i.e., no packet is generated in interval $(g_{\ell (i-1)}^\pi+\av{T}_{\ell}^\pi-\epsilon,g_{\ell (i-1)}^\pi+\av{T}_{\ell}^\pi+\epsilon)$ is $\bbP(\epsilon_{\ell i}^2>\epsilon^2)=e^{-(2\epsilon)/\mu_\ell}$, $\forall i$. Therefore, using strong law of large numbers \cite{patrick1995probability}, with probability 1,
\begin{align} \label{eq:beta-lb-2}
\lim_{t\to\infty}\frac{\sum_{i=1}^{R_\ell^\pi(t)}\epsilon_{\ell i}^2}{R_\ell^\pi(t)}=\bbE[\epsilon_{\ell i}^2]
&=\int_{0}^\infty \bbP(\epsilon_{\ell i}^2> \epsilon^2)d\epsilon^2, \nonumber \\
&=\int_{0}^\infty e^{-(2\epsilon)/\mu_\ell} d\epsilon^2, \nonumber \\ 
&=\int_{0}^\infty 2\epsilon e^{-2\epsilon/\mu_\ell} d\epsilon, \nonumber \\ 
&=\frac{\mu_\ell^2}{2}. 
\end{align}

From \eqref{eq:beta-lb-1} and \eqref{eq:beta-lb-2}, we get Proposition \ref{prop:lb-beta}.

\section{Proof of Proposition \ref{prop:Tav-beta-M}} \label{Appendix-proof-prop-Tav-beta-M}

\subsection{$\av{T}_\ell^M=\bbE_M[T_{\ell i}^M]=\av{T}_\ell^{\max}$.}
\begin{IEEEproof}
Recall that the packets in $\cM_\ell$ are marked using a threshold policy $\pi_{Th}$, with threshold $A_\ell=\av{T}_\ell^{\max}-\mu_\ell$. Therefore, $T_{\ell i}^M=(\av{T}_\ell^{\max}-\mu_\ell)+X_\ell$, where $X_\ell$ is an exponentially distributed random variable with mean $\mu_\ell$ that denotes the earliest time instant (relative to the instant when the time elapsed since the generation time of previously marked packet equals $A_\ell$) when a packet at source $\ell$ gets generated.
Thus, as shown in \cite{saurav2021minimizing}, $\bbE_M[T_{\ell i}^M]=\av{T}_{\ell}^{\max}$. Also, $\pi_{Th}$ is a stationary policy, for which $T_{\ell i}^M$, $\forall i\ge 1$, are independent and identically distributed (shown in \cite{saurav2021minimizing}). Hence, using strong law of large numbers (Remark \ref{remark:RM-to-infty} below),  
$\av{T}_{\ell}^M=\bbE_M[T_{\ell i}^M]=\av{T}_{\ell}^{max}$.
\end{IEEEproof}

\begin{remark} \label{remark:RM-to-infty}
From Remark \ref{remark:Tmax-finite}, we know that $\av{T}_{\ell}^{max}$ is finite. Also, $\bbE_M[X_\ell]=\mu_\ell<\infty$ (inter-generation time of packets at source $\ell$ is exponentially distributed with finite mean $\mu_\ell$, independent of the policy). Hence, $\bbE_M[T_{\ell i}^M]$ is also finite ($T_{\ell i}^M$ ($\forall i\ge1$) are finite with probability 1). Hence, as $t\to\infty$, the number of packets in $\cM_\ell$, i.e., $R_\ell^M(t)$ (that we consider as transmitted), also approaches infinity. Hence, using the fact that $T_{\ell i}^M$ ($\forall i\ge1$) are independent and identically distributed, we can use strong law of large numbers \cite{patrick1995probability} to claim that the average of $T_{\ell i}^M$ (for $i=1,\cdots,R_\ell^M(t)$) is $\bbE_M[T_{\ell i}^M]$.
\end{remark}

\subsection{$\beta_\ell^M=\mu_\ell^2$.}
\begin{IEEEproof}
Using strong law of large numbers (Remark \ref{remark:RM-to-infty}), with probability 1, we get 
\begin{align}
    \beta_\ell^M&=\lim_{t\to\infty}\frac{\sum_{i=1}^{R_{\ell}^M(t)}(T_{\ell i}^M-\av{T}_\ell^M)^2}{R_\ell^M(t)}, \nonumber \\
    &=\lim_{t\to\infty}\frac{\sum_{i=1}^{R_{\ell}^M(t)}(T_{\ell i}^M-\bbE[T_{\ell i}^M])^2}{R_\ell^M(t)}, \nonumber \\
    &\stackrel{(a)}{=}Var(T_{\ell i}^M), \nonumber \\
    &=Var(\av{T}_\ell^{\max})+Var(X_\ell), \nonumber \\
    &\stackrel{(b)}{=}\mu_\ell^2,
\end{align}
where in $(a)$,
$Var(T_{\ell i}^M)$ denotes the variance of $T_{\ell i}^M$, and we get $(b)$ because $\av{T}_\ell^{\max}$ is a constant, and hence, its variance is $0$. 
\end{IEEEproof}


\subsection{$\bbE_M[Z_{\ell i}^M]=\bbE_M[W_{\ell i}^M]+\bbE_M[d_{\ell i}]\le \av{T}_\ell^{\max}+\gamma_\ell$.}
\begin{IEEEproof}
Note that $\pi_M$ initiates transmission of packet $\ell_i^M$ at the earliest time instant after $g_{\ell i}^M$, when $\pi_R$ picks source $\ell$ to transmit. Hence, if $t_{\ell (j-1)}^R$ and $t_{\ell j}^R$ denote the successive time instants when $\pi_R$ picks source $\ell$ to transmit, and $g_{\ell i}^M\in(t_{\ell (j-1)}^R,t_{\ell j}^R]$, then $s_{\ell i}^M=t_{\ell j}^R$, and hence, as shown in Figure \ref{fig:R-th-policy},
$W_{\ell i}^M=s_{\ell i}^M-g_{\ell i}^M= t_{\ell j}^R-g_{\ell i}^M\le t_{\ell j}^R-t_{\ell (j-1)}^R$. Thus, 
\begin{align} \label{eq:W-less-Y}
    W_{\ell i}^M\le Y_{\ell j}^R,
\end{align}
where $Y_{\ell j}^R=t_{\ell j}^R-t_{\ell (j-1)}^R$ denotes the time between successive time instants, when $\pi_R$ picks source $\ell$ to transmit. 
 \begin{figure} 
	\begin{center}
		\begin{tikzpicture}[thick,scale=0.8, every node/.style={scale=1}]
		\draw[->] (-0.25,0) to (7.8,0) node[below]{$t$};
		
		\draw (0,-0.1) node[below]{$g_{\ell (i-2)}^M$} to (0,0.1);
		\draw (1.4,-0.1) node[below]{$t_{\ell (j-1)}^R$} to (1.4,0.1);
		\draw (2.9,-0.1) node[below]{$g_{\ell (i-1)}^M$} to (2.9,0.1);
		\draw (4,-0.1) node[below]{$g_{\ell i}^M$} to (4,0.1);
        \draw (5.3,-0.1) node[below]{$t_{\ell j}^R$} to (5.3,0.1); 
        \draw (7,-0.1) to (7,0.1);

		\draw[|<->|] (4,0.4) -- (5.3,0.4) node[rectangle,inner sep=-1pt,midway,fill=white]{$W_{\ell i}^M$}; 
		\draw[<->|] (5.3,0.4) -- (7,0.4) node[rectangle,inner sep=-1pt,midway,fill=white]{$d_{\ell i}^M$};
		\draw[|<->|] (2.9,1.1) -- (5.3,1.1) node[rectangle,inner sep=-1pt,midway,fill=white]{$W_{\ell (i-1)}^M$};
		\draw[<->|] (5.3,1.1) -- (7,1.1) node[rectangle,inner sep=-1pt,midway,fill=white]{$d_{\ell (i-1)}^M$};
		\draw[|<->|] (1.4,1.8) -- (5.3,1.8) node[rectangle,inner sep=-1pt,midway,fill=white]{$Y_{\ell j}^R$};
		
		\draw[|<->|] (0,-1.3) -- (2.9,-1.3) node[rectangle,inner sep=-1pt,midway,fill=white]{$T_{\ell (i-1)}^M$}; 
		\draw[<->|] (2.9,-1.3) -- (4,-1.3) node[rectangle,inner sep=-1pt,midway,fill=white]{$T_{\ell i}^M$};

		\end{tikzpicture}
		\caption{Relationship between the different quantities defined with respect to policy $\pi_M$.\vspace{-2ex}}  
		\label{fig:R-th-policy} 
	\end{center}
\end{figure}
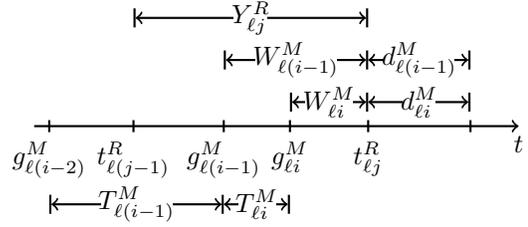

Recall policy $\pi_R$ (Algorithm \ref{algo:threshold-policy}). Each time $\pi_R$ picks a source $\ell$, it waits for $d_\ell\sim\cD_\ell$ time units before picking the next source (either because source $\ell$ transmits a packet and channel becomes busy, or because $\pi_R$ idles due to unavailability of packet at source $\ell$ for transmission). Hence, $Y_{\ell j}^R=\sum_{n=1}^N\sum_{k=1}^{K_{\ell j}(n)}d_{n k}$, where $K_{\ell j}(n)$ denotes the number of times source $n$ is picked by $\pi_R$ in interval $(t_{\ell (j-1)}^R,t_{\ell j}^R]$, and $d_{n k}\sim\cD_\ell$ are independent and identically distributed random variables $\forall k\ge 1$. 

Since $\pi_R$ picks each source with fixed probability, we know that $K_{\ell j}(n)$ is geometrically distributed random variables with mean $p_n/p_\ell$, and $K_{\ell j}(n)$ and $d_{n k}$ are mutually independent, $\forall n,k$. Hence, using Wald's equation \cite{mckay2019probability}, we get $\bbE_M[Y_{\ell j}^R]$. 
\begin{align} \label{eq:EY-1}
    \bbE_R[Y_{\ell j}^R]=\sum_{n=1}^N(\bbE_R[K_{\ell j}(n)]\bbE_R[d_{n k}])=\sum_{n=1}^N \frac{p_n}{p_\ell}\gamma_n.
\end{align}
Substituting for $p_n$ and $p_\ell$ from \eqref{eq:pick-prob} into \eqref{eq:EY-1}, we get
\begin{align} \label{eq:EY-less-Tmax}
    \bbE_R[Y_{\ell j}^R]=\av{T}_\ell^{\max}\sum_{n=1}^N\frac{\gamma_n}{\av{T}_n^{\max}}\stackrel{(a)}{\le}\av{T}_\ell^{\max}, 
\end{align}
where $(a)$ follows from the second property in Lemma \ref{lemma:NC}.
Thus, from \eqref{eq:W-less-Y} and \eqref{eq:EY-less-Tmax}, we get $\bbE_M[W_{\ell i}^M]\le \av{T}_\ell^{\max}$.

Further, from Remark \ref{remark:d-independent-pi}, we know that $d_{\ell i}\sim\cD_\ell$ are independent and identically distributed ($\forall i\ge1$, with mean $\gamma_\ell$), independent of policy. Hence, $\bbE_M[d_{\ell i}]=\gamma_\ell$. 

Since $Z_{\ell i}^M=W_{\ell i}^M+d_{\ell i}$ (by definition), therefore, we get
$\bbE_M[Z_{\ell i}^M]=\bbE_M[W_{\ell i}^M]+\bbE_M[d_{\ell i}]\le \av{T}_\ell^{\max}+\gamma_\ell$.
\end{IEEEproof}

\end{document}